\documentclass{article} 

\usepackage[bindingoffset=0.2in, left=1.in,right=1.in,top=0.75in,bottom=1in, footskip=.25in]{geometry} 

\usepackage{abstract}
\usepackage{authblk}
\usepackage{blindtext}
\usepackage{sectsty}

\sectionfont{\fontsize{12}{15}\selectfont}
\subsectionfont{\fontsize{10}{15}\selectfont}

\usepackage{graphicx}
\usepackage{amsmath}
\usepackage{amssymb} 
\usepackage{upgreek}
\usepackage[normalem]{ulem}
\usepackage{bm} 
\usepackage[utf8]{inputenc} 
\usepackage[american]{babel} 
\usepackage[bookmarks=false]{hyperref}
\hypersetup{colorlinks={true},linkcolor={blue},citecolor={blue},urlcolor={blue}}
\usepackage{lscape}

\usepackage{siunitx}

\usepackage{textcomp,gensymb}

\usepackage[square]{natbib}

\usepackage{enumitem}

\newcommand{\D}{\Delta}
\newcommand{\quotes}[1]{``{#1}"}
\newcommand{\Ot}{\widehat{O}}
\newcommand{\xt}{\widehat{x}}
\newcommand{\zt}{\widehat{z}}
\newcommand{\ut}{\widehat{u}}

\newcommand{\pd}[2]{\partial_{#2}{#1}} 
\newcommand{\td}[2]{\frac{\textrm{d}#1}{\textrm{d}#2}} 
\newcommand{\boldnabla}{\boldsymbol{\nabla}} 

\newcommand{\modif}[1]{#1}
\newcommand{\modtwo}[1]{#1}
\newcommand{\modthree}[1]{#1}

\def \reff#1 {
	(\ref{#1})
}

\usepackage{soul}
\usepackage[usenames, dvipsnames]{color} 

\usepackage{float}
\usepackage{placeins}
\usepackage{caption}

\def \fig [#1][#2]#3#4 {
  \begin{figure}
    \centering
      \includegraphics[width=#2\textwidth]{figs/#1}
      \caption{#3}{\footnotesize{#4}}
      \label{fig:#1}
  \end{figure}
}

\def \subfigs [#1]#2#3#4 {
  \begin{figure}
    \centering
      #2
      \caption{#3}{\footnotesize{#4}}
      \label{fig:#1}
  \end{figure}
}

\def \subfig [#1][#2]#3 {
  \begin{subfigure}[b]{#2\textwidth}
    \includegraphics[width=\textwidth]{figs/#1}
    \caption{#3}
    \label{fig:#1}
  \end{subfigure}
}

\usepackage{etoolbox}

\makeatletter

\patchcmd{\NAT@citex}
  {\@citea\NAT@hyper@{%
     \NAT@nmfmt{\NAT@nm}%
     \hyper@natlinkbreak{\NAT@aysep\NAT@spacechar}{\@citeb\@extra@b@citeb}%
     \NAT@date}}
  {\@citea\NAT@nmfmt{\NAT@nm}%
   \NAT@aysep\NAT@spacechar\NAT@hyper@{\NAT@date}}{}{}

\patchcmd{\NAT@citex}
  {\@citea\NAT@hyper@{%
     \NAT@nmfmt{\NAT@nm}%
     \hyper@natlinkbreak{\NAT@spacechar\NAT@@open\if*#1*\else#1\NAT@spacechar\fi}%
       {\@citeb\@extra@b@citeb}%
     \NAT@date}}
  {\@citea\NAT@nmfmt{\NAT@nm}%
   \NAT@spacechar\NAT@@open\if*#1*\else#1\NAT@spacechar\fi\NAT@hyper@{\NAT@date}}
  {}{}
  
\makeatother

\usepackage{color}

\title{\bf Internal wave boluses as coherent structures\\ in a continuously stratified fluid}

\author{Guilherme S. Vieira\thanks{salvadorvieira.g@northeastern.edu} }
\author{Michael R. Allshouse\thanks{m.allshouse@northeastern.edu}}

\affil{\normalsize{Department of Mechanical and Industrial Engineering, Northeastern University, \\}
\normalsize{Boston, MA 02115, USA\\}

\vspace{1em}
(Dated: 22 November 2019)
}

\date{}

\begin{document}

\renewcommand{\abstractname}{}    
\renewcommand{\absnamepos}{empty} 
 
\maketitle

\vspace{-2em}


\begin{abstract}

\begin{center}
    \textbf{Abstract}
\end{center}
\vspace{0.5em}

Internal waves shoaling on the continental slope can break and form materially coherent vortices called boluses.  These boluses are able to trap and transport material up the continental slope, yet the global extent of bolus transport is unknown.  Previous studies of bolus formation primarily focused on systems consisting of two layers of uniform density, which do not account for the presence of ocean pycnoclines of finite thickness. We use hyperbolic tangent profiles to model the density stratification in our simulations and demonstrate the impact of the pycnocline on the bolus.  A spectral clustering method is used to objectively identify the bolus as a Lagrangian coherent structure that contains the material advected upslope.  The bolus size and displacement upslope are examined as a function of the pycnocline thickness, incoming wave energy, density change across the pycnocline, and topographic slope.  The dependence of bolus transport on the pycnocline thickness demonstrates that boluses in continuous stratifications tend to be larger and transport material further than in corresponding two-layer stratifications.

\end{abstract}



\section{Introduction} \label{sec:Intro}


Temperature and salinity variations stratify the oceans, which enables the propagation of density disturbances as internal waves.
These waves are often generated by tides passing over topography or as a result of surface storms~\citep{alford03,wunsch04} and play an important role in transferring energy and momentum throughout the ocean~\citep{munk98}.
Internal waves are particularly prominent in the pycnocline, the region of sharp transition between the upper mixed layer and the deep ocean, where density gradients are strongest.
These waves have amplitudes ranging from tens to hundreds of meters~\citep{duda04,susanto05,helfrich06} and can travel hundreds to thousands of kilometers from their sources~\citep{osborne80,ray96} before breaking on the continental slope~\citep{troy05,lamb14}.  
The turbulence resulting from internal wave breaking plays an important role in ocean mixing and energy dissipation~\citep{sandstrom95,inall00,moum03}, but the transport induced is not fully known. 


The propagation of internal waves above the continental slope towards the coastline causes these waves to undergo shoaling, a process associated with an increase in amplitude that ultimately \modif{results in} the breaking of the wave.  
This process can result in the formation of a bolus, a moving vortex capable of transporting water and suspended particulate up the slope~\citep{helfrich86,helfrich92,venayagamoorthy07, fructus09,aghsaee10,lamb14}. 
Boluses have been observed in the ocean~\citep{carter05,moum07,walter12,alford15} and can span half the water column height~\citep{klymak03}.

It is not known whether boluses transport a significant amount of bio-matter or sediments up the continental slope globally, but it has been proposed that the upwelling and turbulent mixing supported by this phenomenon could be vital for transporting nutrient-rich fluid into coastal ecosystems~\citep{klymak03,wang07}.  On the Scotian Shelf, for example, mixing resulting from internal tide breaking is believed to act as a nutrient pump from the deeper waters to the euphotic zone~\citep{sandstrom84}.  
\modtwo{
Boluses formed by internal tides interacting with topography are also one of the best known larval transport mechanisms in open coastline populations. The characteristic advection of water masses and development of warm-water fronts are key to the onshore transport of larvae~\citep{pineda91,pineda94}.}

It has also been hypothesized that shoaling internal waves are effective agents for sediment resuspension because the wave destabilization and subsequent turbulence drive sediments out of the bottom boundary layer and further up into the water column\modtwo{~\citep{hosegood04,quaresma07,stastna08, bourgault14, boegman19}}, and strong turbulent mixing and sediment resuspension associated \modtwo{ with} the run-up of internal wave boluses have been observed \modthree{in Otsuchi Bay}~\citep{masunaga15}.
An estimate of the influence of boluses on the transport of nutrients or their impact on resuspension, however, remains to be determined.


To better understand the breaking process and potential transport by boluses, laboratory experiments have been conducted to study shoaling internal waves on a slope~\citep{cacchione74, dauxois04} and bolus dynamics in approximately two-layer stratified systems~\citep{helfrich92, michallet99, boegman05, sutherland13, moore16}.  
The two-layer systems used in these experiments correspond to an idealized ocean stratification. 
In the two-layer system, solitary waves or wave trains propagate \modif{along the pycnocline} and shoal onto a constant slope topography, potentially generating boluses.
\citet{helfrich92} investigated the interaction of a solitary wave of depression with a sloping bottom and described the break up of the incoming wave into several boluses.
Using a similar system, \citet{michallet99} described the breaking of large-amplitude internal waves, identifying the first sign of breaking as a gravitational instability. Using particle image velocimetry, they visualized the vortex created \modtwo{at the breaking location} and suggested that this mechanism could sweep material originally close to the slope far offshore.  


While experiments provide reliable measurements and allow for a description of the physics of the breaking process and bolus propagation, numerical simulations furnish a more complete representation of the bolus, the breaking mechanism and their characteristics~\citep{lamb03,legg03,venayagamoorthy06, venayagamoorthy07,aghsaee10, arthur14, arthur16, arthur17}.
Two- and three-dimensional laboratory-scale numerical simulations in a linearly stratified system were performed by \citet{venayagamoorthy07}, to study the formation and propagation of nonlinear boluses \modtwo{produced at a shelf break}.
They demonstrated that the three-dimensional bolus structure is stable in the transverse dimension, therefore justifying the use of two-dimensional simulations as an accurate representation of the bolus dynamics.
More recently, \citet{arthur16} investigated transport due to breaking internal waves on slopes by incorporating particle-tracking \modthree{in} three-dimensional simulations, demonstrating onshore and offshore transport within the bolus, as well as lateral particle transport away from the topography due to turbulence developed by the breaking. 
\modif{\citet{masunaga17} modeled sediment resuspension due to boluses with a transport equation for suspended sediment, obtaining resuspension processes that are in good agreement with observational data and investigating the effect of varying the topographic slope.}
\citet{fringer03} and \citet{arthur17} departed from the two-layer stratification model to investigate the impact of a finite pycnocline thickness on mixing, dissipation and turbulence due to breaking internal waves on a slope.

\begin{figure}
  \centerline{\includegraphics[width=1.22\textwidth]{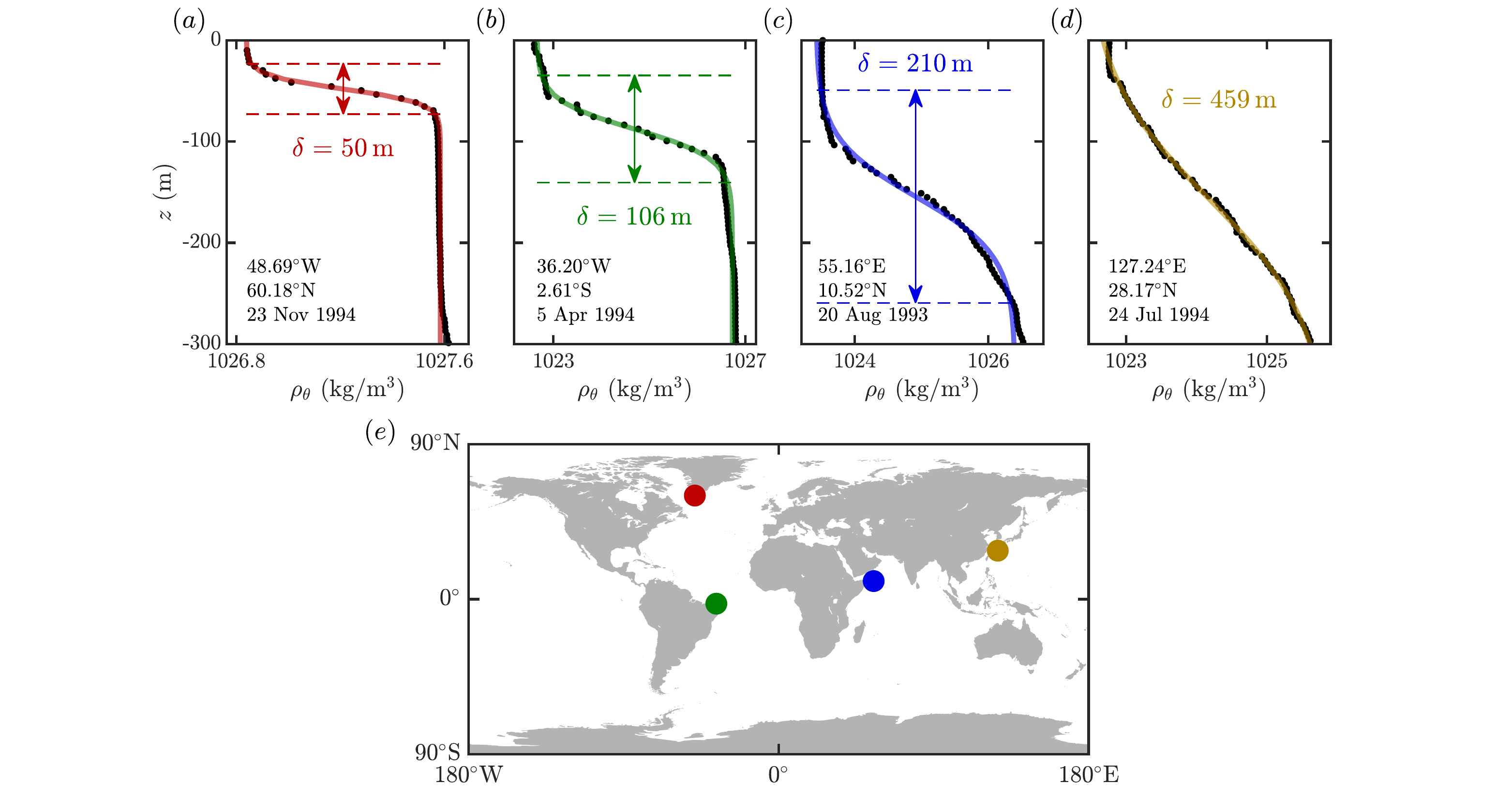}}
  \caption{Potential density profiles $\rho_\theta(z)$ near the continental shelf obtained from the World Ocean Circulation Experiment Hydrographic Program~\citep{schlitzer00}. $(a)$-$(d)$ Observational data (black dots) and the best fit line using a hyperbolic tangent function (colored line).  The pycnocline thickness $\delta$ describes the width of the transition in density. $(e)$ The geolocation of each density profile is identified with a marker corresponding to the color of the best fit line: $(a)$ red, $(b)$ green, $(c)$ blue, and $(d)$ yellow.}
\label{fig:fig1}
\end{figure}

No previous work on bolus characterization, however, has analyzed the impact of the pycnocline thickness on the generation of boluses and the resulting transport.
From salinity and temperature measurements taken during the World Ocean Circulation Experiment at over 18,000 stations worldwide~\citep{schlitzer00}, it is evident that an improved description of the upper ocean stratification can be achieved by using a hyperbolic tangent profile~\citep{maderich01} and parameterization of the pycnocline thickness $\delta$.
The limiting cases of high and low $\delta$ correspond to the linearly stratified and two-layer density models, respectively. 
Figure~\ref{fig:fig1} presents the potential density profiles at four stations within \SI{300}{km} of the coastline, where boluses are prone to form.
The presented profiles have values of $\delta$ ranging from $\SI{50}{m}$ in the Labrador Sea, south of Greenland (figure~\ref{fig:fig1}$a$), to $\SI{459}{m}$ in the East China Sea (figure~\ref{fig:fig1}$d$), where the profile is nearly linear.
Even for the thinnest pycnocline, the two-layer density model does not provide a close approximation.
These examples highlight the fact that stratification profiles, and in particular the pycnocline thicknesses, vary worldwide with the geographical location and the season~\citep{aikman84,liu01,sigman04}, and such variation may have an impact on the formation and propagation of boluses.


Another essential approach for understanding the impact of boluses is the quantification of bolus transport characteristics, which requires an objective definition of the bolus itself.
In two-layer systems, the bolus can be naturally visualized as a propagating front of denser fluid, but such a definition is not straightforward in continuously stratified systems. For a linear stratification, \citet{venayagamoorthy07} propose two ways to define the bolus speed from the density field: the speed at which the bolus front travels as observed via isopycnals, or the speed minimizing the time rate of change of the density field in a co-moving reference frame. 
Both methods describe the bolus dynamics from an Eulerian perspective, directly from the density field, and therefore do not necessarily represent transport of fluid elements.
\citet{arthur16} more accurately quantify transport by incorporating a particle-tracking model to the simulations, but the question of how much of the transport is due to boluses or to the general breaking dynamics remains undetermined.


In this work, we identify boluses as elliptic Lagrangian coherent structures~\citep{froyland15, allshouse15}, which are regions of the fluid that do not significantly mix with the rest of the domain. 
These objectively defined structures identify materially coherent vortices in the Lagrangian frame~\citep{haller13,haller16,serra17}. 
This characterization provides a precise description of the phenomenon from a transport perspective because it exclusively captures the material transported by the bolus. 
We identify boluses by applying a clustering algorithm to the trajectories of passive tracers that are advected by the breaking dynamics. This approach is based on the method presented by \citet{hadjighasem16}, which can identify vortex-like coherent structures.

We investigate in this paper the impact of the pycnocline thickness on the dynamics and transport properties of internal wave boluses.  
We use coherent structures to quantify the transport properties of boluses, and our numerical simulations demonstrate the dependence of bolus transport on the pycnocline thickness, the incoming wave energy, the density change in the pycnocline, and the topographic slope. 
The computational approach and a sample simulation of a bolus forming as the internal wave breaks on a constant slope is presented in~\S\ref{sec:Numerical}. 
The characterization of the bolus from the Lagrangian coherent structure perspective, the transport metrics, \modif{and a comparison of the results between two- and three-dimensional models} are presented in~\S\ref{sec:Boluses}. 
The dependence of bolus characteristics \modtwo{on} the stratification, wave properties, topography and relevant dimensionless parameters is presented in~\S\ref{sec:Parametric}.  
Finally, the conclusions, potential applications and possible extensions of this work are discussed in \S\ref{sec:Conclusions}.

\section{Numerical model} \label{sec:Numerical}


This section discusses the numerical simulations of the internal wave breaking on a constant slope and the resulting formation and propagation of boluses.  
In \S\ref{sec:computational_approach}, the governing equations, numerical domain, system forcing, relevant parameters and measured quantities are presented.  
The breaking dynamics and bolus propagation up the slope for a sample simulation are illustrated in \S\ref{sec:density_perturbation} from an Eulerian perspective. 

\subsection{Computational approach, domain and setup} \label{sec:computational_approach}


The Navier-Stokes equations in an inertial frame, using the Boussinesq approximation for an inhomogeneous, incompressible fluid subject to gravity along the vertical direction $z$ are used to simulate the laboratory-scale system.  
The stability of the bolus structure in the transverse direction as it propagates up the slope~\citep{venayagamoorthy07} gives credence to using the two-dimensional model.  \modif{However, we will verify \modthree{that} the coherent structure analysis of a three-dimensional simulation produces a similar bolus to the two-dimensional case in \S\ref{sec:2d_3d}.} 
The Boussinesq approximation neglects effects of density variation except in the buoyancy term, and is appropriate for buoyancy-driven flows with weak relative density variations around a reference value $\rho_{00}$. 
This approximation has been extensively used in ocean models and in two-layer density systems with small relative density change~\citep{long65,helfrich06,pedlosky13}. The system of equations is
\begin{eqnarray}
\label{eq:NS-1}
\boldnabla \cdot \mathbf{u} &=& 0, \\
\label{eq:NS-2}
\pd{u}{t} + (\mathbf{u} \cdot \boldnabla) u
 &=& -\frac{1}{\rho_{00}}\pd{p}{x} + \nu \boldnabla^2 u, \\
\label{eq:NS-3}
\pd{w}{t} + (\mathbf{u} \cdot \boldnabla) w
 &=& -\frac{1}{\rho_{00}}\pd{p}{z} + \nu \boldnabla^2 w - \frac{\rho}{\rho_{00}}g, \\
\label{eq:NS-4}
\pd{\rho}{t} + (\mathbf{u} \cdot \boldnabla) \rho & = & \kappa \boldnabla^2\rho,
\end{eqnarray}
where $\mathbf{u} = (u,w)$ is the velocity field, $p$ is the pressure, $\rho$ is the local density, $\rho_{00} = \SI{1000}{kg/m^{3}}$ is the reference density, $\nu =\SI{e-6}{m^2/s}$ is the kinematic viscosity, $\kappa = \SI{1.4e-7}{m^2/s}$ is the thermal diffusivity for sea water~\citep{kunze03}, and $g = \SI{9.8}{m/s^{2}}$ is the gravity acceleration. 
It is assumed in \eqref{eq:NS-4} that density diffusion in the system is driven by thermal diffusion.


The system of equations \eqref{eq:NS-1}-\eqref{eq:NS-4} is solved numerically for $u$, $w$, $p$ and $\rho$ using CDP~2.4, an unstructured, finite-volume based, large eddy simulation code that implements a fractional-step time-marching scheme~\citep{ham04, mahesh04}.  
All subgrid‐scale modelings are turned off, and the code is modified to include the buoyancy term in \eqref{eq:NS-3} and solve \eqref{eq:NS-4} along with \eqref{eq:NS-1}-\eqref{eq:NS-3}.  
This code has previously been used to simulate internal waves and has been validated with experiments~\citep{king09,dettner13,lee14,paoletti14,zhang14,allshouse16, lee18}.

\begin{figure}
  \centerline{
  \includegraphics[width=1.2\textwidth]{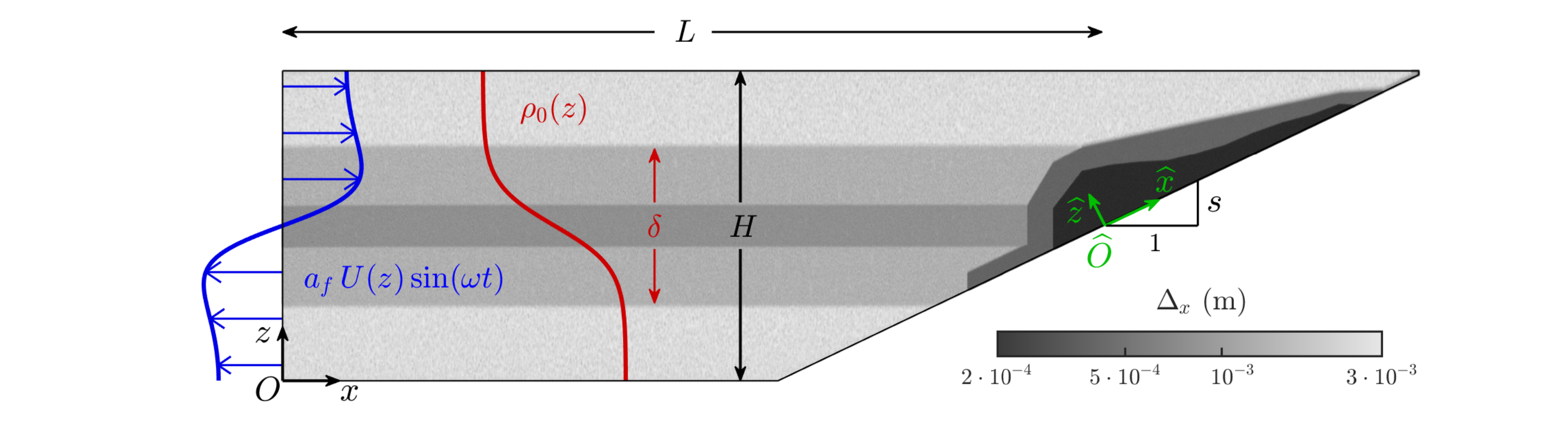}
  }
  \caption{Schematic diagram of the computational domain, dimensions, mesh resolution, density profile (in red) and fundamental mode forcing velocity profile (in blue). The distance from the inlet boundary on the left to the midpoint of the slope $L=\SI{2.87}{m}$ and the domain height $H=\SI{0.4}{m}$ are constant in all simulations, while $\delta$ and $s$ are varied in the parametric study. A sample background density profile $\rho_0(z)$ (red) and inlet velocity profile $U(z)$ (blue) are presented corresponding to the sample simulation ($\delta=\SI{0.2}{m}$,  $\D\rho=\SI{20}{kg/m^3}$, $s=0.176$, presented in~\S\ref{sec:density_perturbation}). The mesh resolution of the sample simulation is presented in a gray logarithmic scale with the highest resolution $\Delta_x=\SI{2e-4}{m}$ in the breaking region (black) and the lowest resolution $\SI{3e-3}{m}$ in the evanescent region (light gray). 
  }
\label{fig:fig2}
\end{figure}

An illustration of the domain is presented in figure~\ref{fig:fig2}.  
The system dimensions correspond to an available system used for experimental studies~\citep{allshouse16}, with similar length scales as in previous experimental~\citep{sutherland13,moore16} and numerical~\citep{venayagamoorthy07,arthur14} bolus investigations.  
The primary reference frame $O_{xz}$ is positioned at the left-bottom corner of the domain. 
At the left-end, a wave-maker generates an internal wave that propagates towards the slope. 
A constant slope topography is positioned at the right-end of the system. 
The \modif{topography, with constant slope $s$,} is positioned such that the mid-depth line ($z=H/2$) intersects the slope at a distance $L = \SI{2.87}{m}$ from the left boundary. 
This point on the slope is used as the origin for the rotated reference frame $\Ot_{\xt\zt}$, where $\xt$ and $\zt$ represent the coordinates along and normal to the slope, respectively.  \modif{Because the location of $\Ot$ is held constant, there is a minimum slope of $s=0.070$  below which the topography would reach the inlet.  For there to be a constant depth development region, the minimum slope studied here is $0.105$.}

In order to efficiently obtain the necessary accuracy for a direct simulation, unstructured triangular multi-block meshes with different resolution zones were adopted. 
The variable mesh resolution is illustrated in figure~\ref{fig:fig2}. 
\modtwo{While the mesh resolution is lower in weakly stratified regions, higher resolutions are used in regions of internal wave propagation ($z\in[0.1, 0.3]\,$m), and the highest resolution is used in the overturning and breaking region on slope.} For the representative mesh in figure~\ref{fig:fig2}, the mesh size, defined as the local averaged distance between cell centers, varies from $\Delta_x=\SI{3e-3}{m}$ in the sparse regions, to $\SI{e-3}{m}$ at the internal wave propagation zone, to a maximum resolution of $\SI{2e-4}{m}$.  
The meshes used for the simulations presented in this work contain between 3.8 and 5.5 million cells. Approximately $70\%$ of these cells are in the breaking zone (corresponding to the darkest zone in figure~\ref{fig:fig2}), where the complex velocity field needs to be accurately resolved.
Simulations with the narrowest pycnocline thickness required a higher resolution mesh to provide converged results, with a total of 12.6 million cells and smallest cell size in the breaking zone reduced to $\SI{5e-5}{m}$.


Convergence studies to define the required spatial and temporal resolutions have been conducted varying both the mesh resolution in the breaking zone and the time step used in the simulations. 
This study resulted in the use of a constant time step of $\SI{5e-4}{s}$ for all simulations.  
With velocity magnitudes below $\SI{0.05}{m/s}$ and the smallest mesh size of $\SI{5e-5}{m}$, the maximum Courant number is~$0.5$. The sea water values for the kinematic viscosity and thermal diffusivity~\citep{kunze03} correspond to a Prandtl number of~$Pr = \nu/\kappa = 7.14$. 
The Reynolds number in the bolus forming region can be estimated using a characteristic length~$L_c = \SI{0.02}{m}$ and a characteristic speed~$U_c = \SI{0.03}{m/s}$ based on the bolus average size and speed, respectively, which corresponds to~$Re = U_c L_c/\nu = 600$. 
Therefore, the Kolmogorov length scale in the breaking region is~$\eta = L_c Re^{-3/4} \approx \SI{1.65e-4}{m}$. 
With a default mesh resolution of $\Delta_x=\SI{2e-4}{m}$ in the breaking zone, of the same order as the microscale $\eta_k$, and $\SI{5e-5}{m}$ for the sharpest pycnocline thickness case, the breaking region is sufficiently resolved to capture the bolus dynamics. 

\begin{figure}
  \centerline{\includegraphics[width=0.7\textwidth]{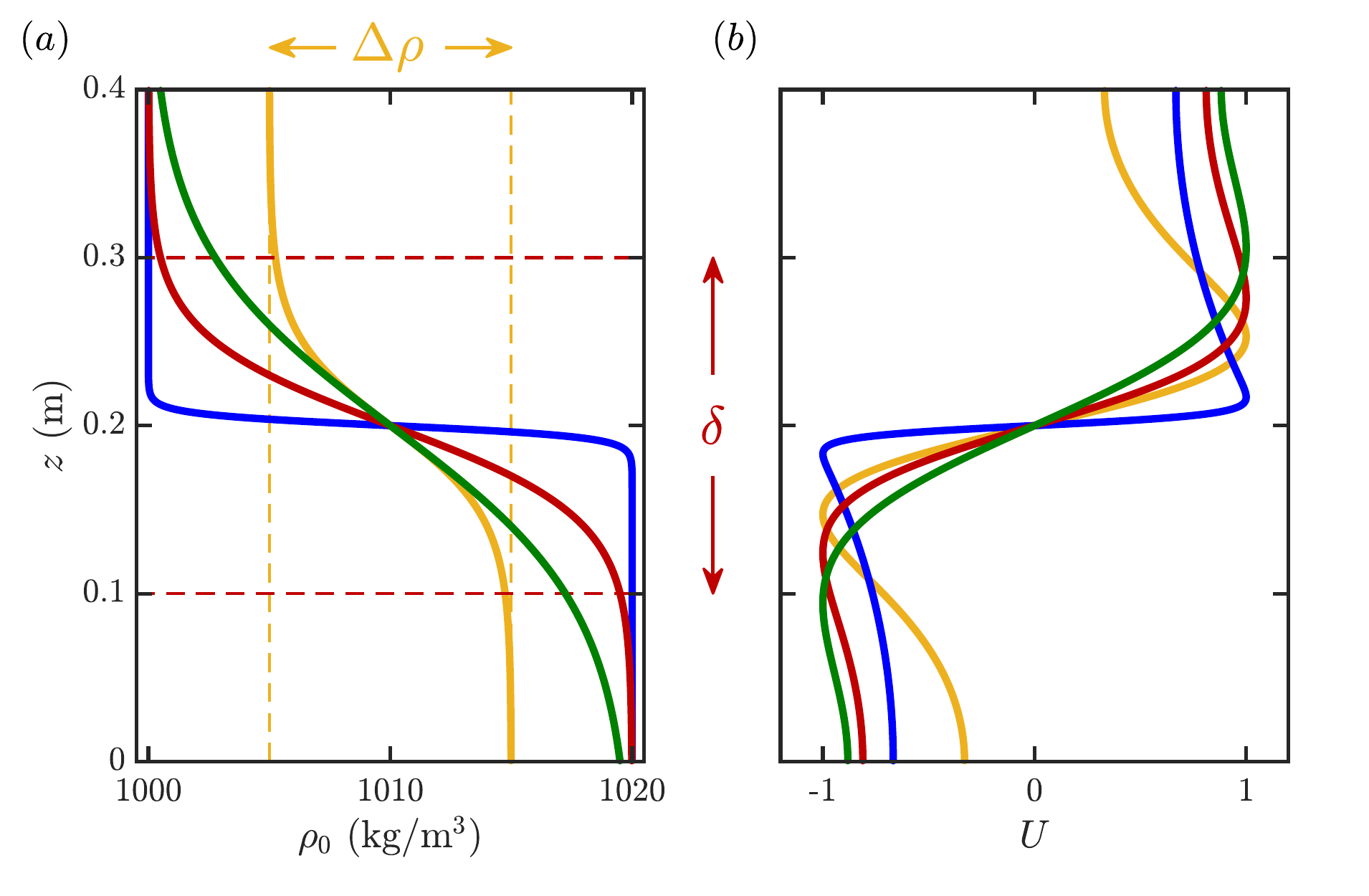}}
  \caption{\modif{$(a)$ Background density stratification for a density change $\D\rho=\SI{20}{kg/m^3}$ and pycnocline thicknesses $\delta=0.025$ (blue), 0.2 (red) and $\SI{0.4}{m}$ (green), as well as $\D\rho=\SI{10}{kg/m^3}$ and $\delta=\SI{0.2}{m}$ (yellow). \modtwo{$(b)$ The normalized} mode-1 velocity profile amplitude $U(z)$ for each corresponding stratification, \modtwo{with frequency $\omega=\SI{0.628}{rad/s}$}.  The red horizontal dashed lines indicate the pycnocline thickness for $\delta=\SI{0.2}{m}$, and the yellow vertical dashed lines the density change for $\D\rho=\SI{10}{kg/m^3}$.}}
\label{fig:fig3}
\end{figure}

The background density stratification is a continuous, decreasing function of $z$. For all simulations, a hyperbolic tangent profile of variable pycnocline thickness is used to provide a smooth transition between the densities at the top and bottom, representative of what is observed in the oceans (figure~\ref{fig:fig1}). The stratification is modeled as
\begin{equation} \label{eq:rho(z)}
\rho_0(z) = \rho_{H/2} - \frac{\Delta\rho}{2}\tanh\left[\frac{2(z-H/2)}{\delta}\tanh^{-1}(0.95)\right],
\end{equation}
for $z \in [0,H],$ where $\D\rho \approx \rho_0(0)-\rho_0(H)$ is the density change and $\rho_{H/2}= \rho_{0}(H/2)$ is the mid-depth unperturbed density value. 
The parameter $\delta$ is the pycnocline thickness for the hyperbolic tangent profile and corresponds to the transition height in which the density varies by $95\%$ of $\Delta\rho$, as illustrated in figure~\ref{fig:fig3}$(a)$. 
The hyperbolic tangent profile shape has been previously used to model sharp density stratifications in the limiting case of an almost two-layer density fluid (corresponding to small $\delta$) because of its stability properties compared to a discontinuous profile~\citep{thorpe71,fringer03,troy05,arthur17}.  \modif{Because the pycnocline is located at the center of the domain, two-layer theory \modthree{does not predict mode-1} internal solitary wave propagation~\citep{long56}.  Unlike~\cite{arthur14}, we will not initialize the wave with a form similar to the solitary wave solution, and instead force the system with a vertical mode-1 profile.}

At $t=0$, the unperturbed system is at rest ($u=w=0$), the density field is $\rho=\rho_0(z)$ and the hydrostatic pressure field is $p=p_0(z)$. As illustrated on the left boundary of the domain in figure~\ref{fig:fig2}, the system is perturbed from its quiescent state by forcing the horizontal velocity $u$ of \modif{the vertical mode-1 plane wave} of frequency $\omega = \SI{0.628}{rad/s}$ (\SI{10}{s}-period). 
The forcing mechanism reproduces numerically what would be obtained experimentally by an oscillating plate wave-maker~\citep{mercier10}.
The \modif{mode-1 wave} is particularly interesting as it is known that internal wave generation mechanisms create waves with the majority of the energy in the lowest modes, with the \modif{first} mode being the one with the largest wavelength and lowest shearing stress, and therefore the least affected by viscous dissipation~\citep{gerkema08}.
\modif{The finite-difference approach for determining the vertical mode-1 profile of frequency $\omega$ (and subsequent modes) for an arbitrary stratification is presented in the Supplementary Material.}

For $\rho_0(z)$ in the form \eqref{eq:rho(z)}, the vertical mode $U(z)$ is a function of $\omega$, $H$, $\D\rho$ and $\delta$.
Figure~\ref{fig:fig3}$(b)$ presents how the fundamental mode profiles $U(z)$ differ for density profiles $\rho_0(z)$ varying the value of $\delta$ \modif{and $\D\rho$}. Note that smoother density profiles correspond to smoother velocity profiles.  Thinner pycnoclines are associated with a stronger velocity shear, $\textrm{d}U/\textrm{d}z$, around $z=H/2$.
The left boundary condition for $u$ is given by
\begin{equation}
u(x=0,z,t) =
\begin{cases}
     \modtwo{a_f\, U(z)\sin(\omega t)} & \textrm{for } t \in [0, 2\pi/\omega],\\
     0 & \textrm{for } t>2\pi/\omega,
\end{cases}
\end{equation}
\modtwo{where $U(z)$ is normalized to unit maximum magnitude and $a_f$ prescribes the amplitude of the forcing.} The system is forced for a single period, because we are only interested in the shoaling of the first, leading wave. After one period, the velocity at the left boundary is set to zero. \modtwo{Only the horizontal velocity is modified from the background value for the forcing.} \modif{Because the system is forced with a mode-1 wave, we are not able to independently modify properties of the wave such as wave speed, amplitude or wavelength.  Additionally, \modtwo{$a_f$ is large enough to cause the propagating wave to be nonlinear}, further complicating the relationship between these properties and the wave forcing parameters.}


Changing the stratification profile modifies the shape of the velocity profile $U(z)$ forced on the left boundary (as illustrated in figure~\ref{fig:fig3}), \modtwo{but it does not prescribe $a_f$}. 
To compare similar waves while varying parameters, we want the kinetic energy of the breaking waves to be held constant, \modtwo{and the forcing amplitude $a_f$ is thus set by the amount of energy present in the resulting breaking wave.}
The wave front kinetic energy through the vertical transect at the horizontal position $x$ and time $t$ is quantified as:
\begin{equation}\label{eq:KE}
E_k(x,t) = \int_{0}^{H}\frac{1}{2}\rho(x,z,t)\left[u^2(x,z,t)+w^2(x,z,t)\right] \textrm{d}z,
\end{equation}
which has units of energy per unit area. 
As the wave travels through the domain, it experiences dissipation and dispersion, changing shape, amplitude and speed even before it arrives at the start of the slope.
Such changes depend on the stratification profile and on the amplitude of the forcing, in such a way that injecting waves at the inlet with the same maximum front kinetic energy $E_k(0,\pi/\omega)$ was found not to be equivalent to obtaining waves of equal front kinetic energy at the breaking \modtwo{location}.
To get the same kinetic energy at the breaking location, preliminary simulations are run in a constant depth channel to determine the \modtwo{forcing velocity amplitude~$a_f$}.
This process guarantees that the generated wave fronts have the same instantaneous kinetic energy at the breaking point $x=L$, at the time $t_c(L)$ when the leading wave crest is at $x=L$.
The value used for the kinetic energy as defined in \eqref{eq:KE} is $E_k(L,t_c(L)) = E_{k,0} = \SI{6.734e-3}{J/m^2}$ for all cases.

The boundary conditions for the top, bottom and sloping boundaries are no-slip for velocity, $u=w=0$, and no-flux for density. 
Imposing a no-slip condition at the top boundary, which corresponds to the water surface, does not significantly impact the dynamics because internal wave perturbations are strongest around mid-depth and decrease exponentially towards top and bottom, so that induced vertical velocities rapidly decay. 
To ensure numerical convergence and complement the inlet boundary condition on the left, a small ($\SI{5e-3}{m}$-high) outlet is added to the top-right of the domain\modtwo{, guaranteeing zero net flux within the domain, as expected from \eqref{eq:NS-1}.}
Studies were performed to ensure that the presence, size, and position of the outlet did not influence the dynamics in the breaking region.

\subsection{Density perturbation field and bolus formation onslope} \label{sec:density_perturbation}

Before analyzing the simulation results from a Lagrangian perspective, we present a typical bolus simulation to establish a basic intuition about the life cycle of the bolus.  
The density perturbation field, $\rho'(x,z,t) = \rho(x,z,t) - \rho_0(z)$, provides an Eulerian description of the wave propagation, breaking, bolus formation and propagation onslope.
This field primarily corresponds to density fluctuations resulting from internal wave propagation, and because the bolus corresponds to higher density fluid moving up the slope, it will be highlighted by this field.  
The sample simulation parameters are $\D\rho=\SI{20}{kg/m^3}$, $\delta=\SI{0.2}{m}$ and $s = 0.176$. 
\modtwo{The mode-1 forcing profile for this case has a velocity amplitude of~$a_f=\SI{0.0126}{m/s}$}, corresponding to oscillating plate displacements up to $\SI{0.02}{m}$ at frequency $\omega = \SI{0.628}{rad/s}$. 
This forcing amplitude produces a front kinetic energy \modif{at the breaking site of} $E_k(L,t_c(L)) = E_{k,0}$.


\begin{figure}
  \centerline{\includegraphics[width=1.15\textwidth]{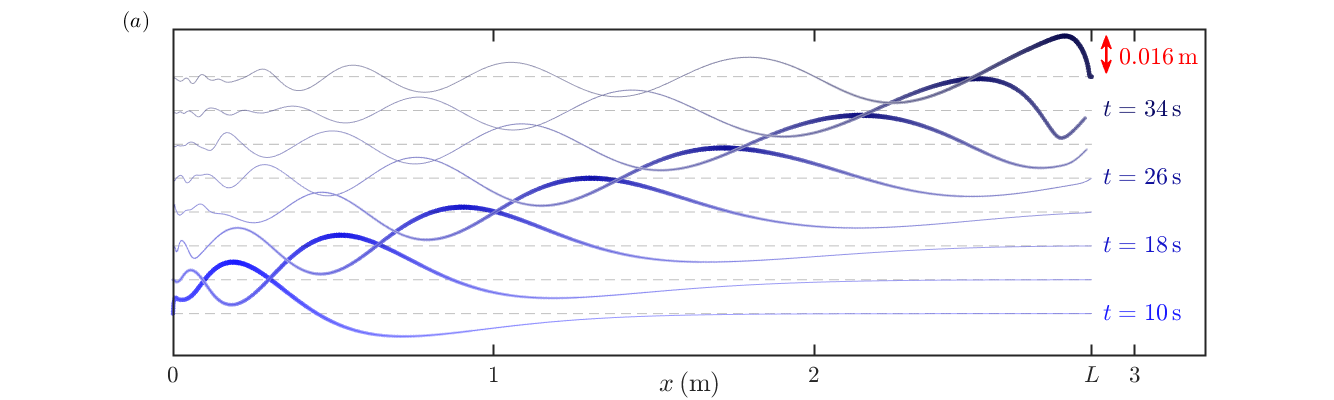}}
  \centerline{\includegraphics[width=1.15\textwidth]{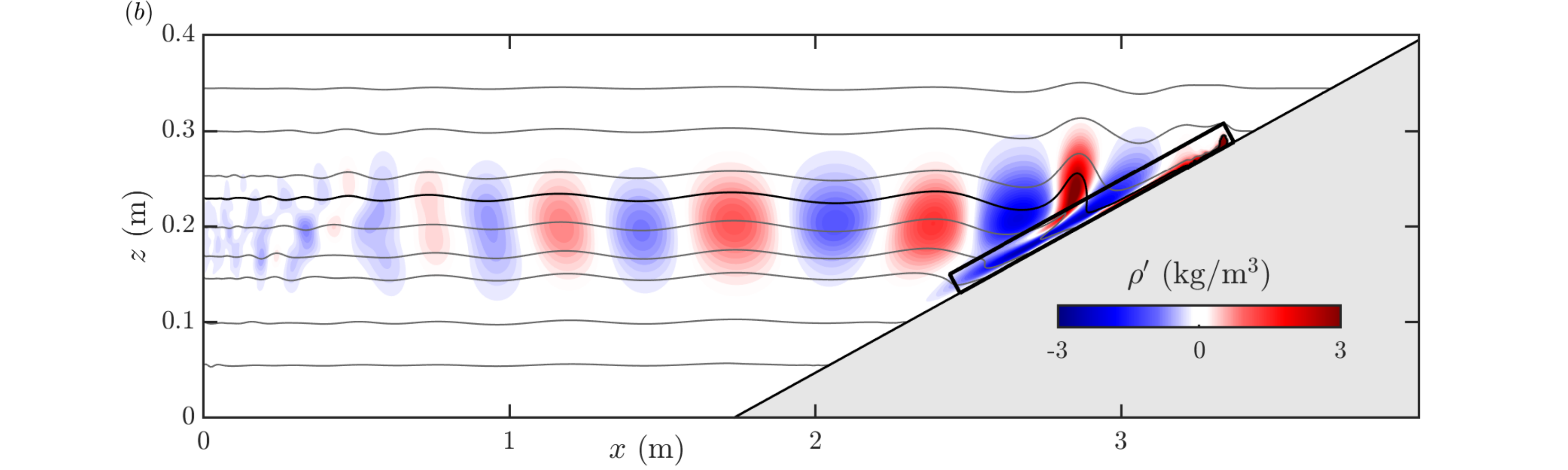}}
  \centerline{\includegraphics[width=1.065\textwidth]{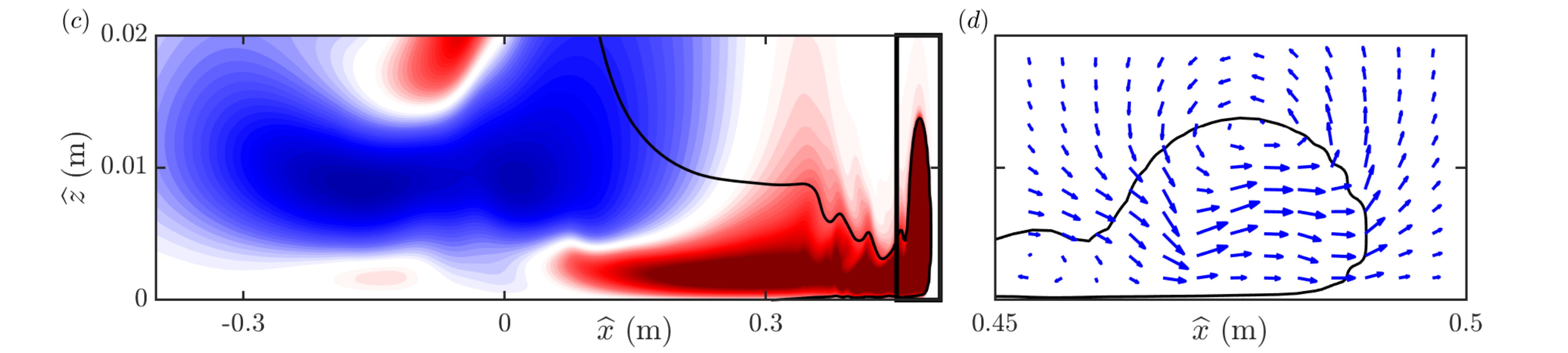}}
  \caption{\modtwo{$(a)$ Time evolution of the mid-depth isopycnal, $\rho_{H/2}=\SI{1010}{kg/m^3}$, with the wave amplitude scale bar indicated in red. The dashed lines represent the unperturbed isopycnal height.} $(b)$ Instantaneous density perturbation field $\rho'$ of the sample simulation \modif{($\delta=\SI{0.2}{m}$)} at $t=\SI{55}{s}$ for the full domain and $(c)$ breaking region. Positive values (red) represent fluid displaced upward and negative values (blue) represent fluid displaced downward.  The solid gray lines in $(b)$ represent isopycnals.
  \modtwo{The isopycnal  $\rho=\SI{1005}{kg/m^3}$ is drawn in black and used to highlight the bolus front boundary in $(c)$ and $(d)$.} The breaking region presented in $(c)$ corresponds to the region surrounded by the thick black box in $(b)$. $(d)$ Velocity field corresponding to the region surrounded by the thick black box in $(c)$, which contains the bolus front.  (Video of the evolution of $(b)$ and $(c)$ is available in the Supplementary Material.)
  }
\label{fig:fig4}
\end{figure}


\modtwo{The wave propagation and the density perturbation field are presented in figure~\ref{fig:fig4}. Figure~\ref{fig:fig4}$(a)$ presents the time evolution of the mid-depth isopycnal $\rho=\rho_{H/2}$, from $t=\SI{10}{s}$, after a full period of forcing, to $t=\SI{38}{s}$, when the isopycnal impinges on the topographic slope at $x=L$. The resulting wave is asymmetric, and its shape slowly changes as it propagates toward the slope.} Note that despite exciting the system for a single period, several \modtwo{higher mode} waves of decreasing amplitude are produced as a result of the initial forcing and persist even after the left boundary conditions are set to zero perturbation (for $t>2\pi/\omega=\SI{10}{s}$). \modtwo{However, these waves move slower than the mode-1 wave and are rapidly left behind and do not affect the bolus dynamics.}

The density perturbation field for the sample simulation at time $t=\SI{55}{s}$ is presented in figure~\ref{fig:fig4}$(b$-$c)$. In this figure, white indicates regions of negligible density perturbation, red indicates positive density perturbation (i.e., denser fluid elements perturbed upwards from their equilibrium position), and blue indicates negative density perturbation (i.e., less dense fluid elements perturbed downwards).
Figure~\ref{fig:fig4}$(b)$ presents the full fluid domain: the generated waves (with alternating crests and troughs indicated in red and blue, respectively) have propagated to the right from the inlet and reached the slope. 
Small asymmetries in the generated waves (the maximum perturbations are not perfectly centered at $z=H/2$) demonstrate that the velocity amplitude is large enough to produce weakly nonlinear waves, even before they reach the sloping region. 
The solid gray lines in \ref{fig:fig4} represent isopycnals, and \modtwo{the isopycnal $\rho=\SI{1005}{kg/m^3}$ (black solid line) highlights the bolus front.}
The negligible density perturbation above and below the pycnocline demonstrates how the \modtwo{density} perturbation amplitudes decrease exponentially in the vertical direction within the weakly stratified, evanescent regions.

The bolus resulting from the internal wave breaking consists of dense fluid moving upslope and is highlighted in figure~\ref{fig:fig4}$(c)$ as a positive (red) density perturbation. 
The results are presented in the rotated frame $\Ot_{\xt\zt}$, and the domain in figure~\ref{fig:fig4}$(c)$ corresponds to the bold, black rectangle ($\SI{0.9}{m}$-long by $\SI{0.02}{m}$-tall) in \ref{fig:fig4}$(b)$. 
The Eulerian results for the bolus propagation are qualitatively similar to those numerically produced by \citet{venayagamoorthy07}.
The internal wave breaking forms the bolus, which corresponds to a propagating counter-clockwise vortex, as illustrated in figure~\ref{fig:fig4}$(d)$ by the velocity field around the bolus front.
The bolus dynamics as it enters the zone of less dense fluid is well described by the theory of gravity currents propagating in a stratified fluid~\citep{benjamin68,maxworthy02,white08}.
The propagation of the dense front entering a zone of lower density fluid is subject to the induced shear and large difference in density between the bolus and the surrounding fluid, which results in vortex shedding and mixing that eventually stops the bolus from moving upslope.
Towards the end of the bolus propagation, the bolus dynamics are impacted by other boluses produced by the following waves reaching the slope. 
These secondary boluses will not be addressed in this manuscript. 
We focus on the dynamics of the first bolus entering a quiescent region with no previous flow that will impact the bolus propagation.


While the density perturbation field and the velocity field provide good intuition for the bolus dynamics, a non-zero value of $\rho'$ at a given point only means that there is no local perturbation at that instant. This approach does not track fluid elements moving with the bolus, nor does this approach provide accurate information about how fluid is transported as the breaking happens.  
This is most acutely highlighted by the fact that the initial internal wave does not transport fluid from the inlet to the sloping boundary.  
At some point during the breaking process, the internal wave does begin to transport fluid (see the Supplementary Material for a demonstration video), and identifying when, how, and what fluid is transported is the purpose of the Lagrangian analysis.

\section{Boluses as Lagrangian coherent structures} \label{sec:Boluses}

To investigate material transport, a Lagrangian approach is more appropriate than an Eulerian approach. This section presents how to use the Eulerian results from the numerical simulations to perform an objective, Lagrangian-based characterization of the bolus, which is here defined as a materially coherent region of the fluid that does not significantly mix with the rest of the domain.
The steps for identifying the bolus, based on the spectral clustering approach developed by \citet{hadjighasem16}, are presented in \S\ref{sec:lagrangian}.  
In \S\ref{sec:bolus_properties}, transport quantities of the bolus are introduced and observations of the relationship between the bolus and the shoaling wave are discussed.
\modif{Finally, a comparison between results for two- and three-dimensional bolus simulations is presented in \S\ref{sec:2d_3d}.}

\subsection{Lagrangian characterization of the bolus}\label{sec:lagrangian}


Lagrangian coherent structures provide a robust means for identifying the key underlying transport features of a given flow~\citep{haller02}.
One type of structure that can be detected is materially coherent vortices~\citep{abernathey18}.  
\modif{These features are unique in that as they move through the domain, the fluid inside the region deforms but does not significantly mix with the fluid outside the perimeter.}  
Identifying this type of feature in the breaking region will determine the fluid that is being advected by the bolus.  
Materially coherent features can be identified using Cauchy-Green based metrics~\citep{haller13}, transport operator methods~\citep{froyland10}, braid-based methods~\citep{allshouse12}, and graph Laplacian methods~\citep{froyland18}.  
We use the clustering based approach based on \citet{hadjighasem16} with minor modifications.

Central to coherent structure detection is the analysis of trajectories, which are computed from the velocity field $\textbf{u}=(u,w)$ obtained by numerically solving  \eqref{eq:NS-1}-\eqref{eq:NS-4}.  The velocity $\mathbf{\ut}$ in the rotated frame is used to advect massless fluid elements referred to as passive tracers, which have no inertia and move according to 
\begin{align}\label{eq:ODE_tracer}
\begin{split}
    \td{\mathbf{\xt}_i}{t} = \mathbf{\ut}(\mathbf{\xt}_i(t),t)
\end{split}
\end{align}
where $\mathbf{\xt}_i(t)$ is the trajectory of tracer $i$ with initial position $\mathbf{\xt}_i(t_0)$.
Numerical integration of \eqref{eq:ODE_tracer} is performed using a 4th-order Runge–Kutta method with a time-step of \SI{5e-3}{s}.
The breaking region is initially covered by a rectangular grid of passive tracers with along-slope dimensions that vary based on the slope and injected energy (grids used for each simulation are in the Supplementary Material).  
\modif{The tracer spacing is $\Delta \xt_p = \SI{5e-3}{m}$ and $\Delta \zt_p = \SI{2e-4}{m}$,} resulting in $n=\,$18,600 to 40,000 tracers.  
The advection time window starts at time \(t_0\), which is just before the shoaling wave arrives at the breaking region, and ends at time \(t_f\), which corresponds to when the initial Eulerian density perturbation has stopped propagating up the slope, an upper bound for when the bolus stopped moving upslope.


Having computed the $n$ trajectories for the interval $[t_0,\,t_f]$, we perform the coherent structure analysis based on the spectral clustering method of~\citet{hadjighasem16}.   
The clustering problem attempts to partition the domain into clusters such that trajectories within the same cluster are similar and trajectories from different clusters are dissimilar.
The spectral analysis relies on the construction of a similarity graph that quantifies the pairwise similarity of trajectories.   The similarity metric \(w_{ij}\) is the inverse of the time-averaged distance between these trajectories.  
This calculation is performed for all pairs of trajectories, but only values of \(w_{ij}\) greater than a user defined value are retained in order to sparsify the similarity matrix, $\mathbf{W}$.
\modif{Here, only values of $w_{ij}>1/r^*$, with $r^*=\SI{0.07}{m}$, are retained, resulting in an approximately $90\%$-sparse matrix for the sample simulation discussed in~\S\ref{sec:density_perturbation}. The clustering results were verified to be consistent for different choices of $r^*$ in a neighborhood of this prescribed value.}
Based on the similarity matrix \modif{$\mathbf{W}$}, the diagonal degree matrix $\mathbf{D}$ is produced, where each diagonal element is equal to the sum of the elements in the corresponding row of \modif{$\mathbf{W}$}. 
From the sparse similarity matrix and degree matrix, the unnormalized graph Laplacian $\mathbf{L} = \mathbf{D} - \mathbf{W}$ is computed.

To create the clustering partition, we must identify characteristics of the set of trajectories. 
To characterize the trajectories, the next step of the algorithm is to compute the first generalized eigenvectors of the generalized eigenproblem
\begin{eqnarray}\label{eq:laplacian_eigenproblem}
\textbf{L}\mathbf{q} = \lambda \textbf{D} \mathbf{q}.
\end{eqnarray}
It has been shown that the generalized eigenvalues of \eqref{eq:laplacian_eigenproblem} satisfy $0=\lambda_1\leq\ldots\leq\lambda_{n}$, and the normalized dominant eigenvectors $\mathbf{q}_1,\mathbf{q}_2,\ldots,\mathbf{q}_n$ differentiate properties in the graph and facilitate the clustering process~\citep{luxburg07}.  
The dominant eigenvectors corresponding to the smallest eigenvalues reveal the most important characteristics of the flow.  
Each trajectory is characterized by a value within each eigenvector, and this characterization is used to group similar trajectories together.  
While all of the eigenvectors provide information, the dominant ones highlight the most important patterns.  


The final step of the spectral clustering algorithm is to construct the eigenvector matrix $\mathbf{Q}~\in~\mathbb{R}^{n\times k}$ containing the $k$ dominant eigenvectors $\mathbf{q}_1,\ldots,\mathbf{q}_k$ as columns. 
While \citet{hadjighasem16} use the eigengap heuristic to determine $k$, this heuristic is not ideal for our application given the small amount of mixing outside of the bolus, so we adopted here a different heuristic based on the form of the dominant eigenvectors to select $k$. $\mathbf{q}_k$ is the first eigenvector highlighting a small fraction of tracers that entrain in a vortex and move upslope.
Let $\mathbf{y}_i \in \mathbb{R}^{k}$ be the characterization vector corresponding to the $i$-th row of $\mathbf{Q}$, which contains condensed differentiating information for trajectory $i$ (note that $k\ll n$). 
The characterization vectors $(\mathbf{y}_i)_{1\leq i \leq n}$ are clustered with a \textit{K}-means algorithm, assigning each vector $\mathbf{y}_i$ and the corresponding trajectory to a cluster.  
We use $k+1$ clusters to partition the domain, where an extra cluster is added to account for the incoherent cluster, as suggested by \citet{hadjighasem16}.

\begin{figure}
  \centerline{\includegraphics[width=1.15\textwidth]{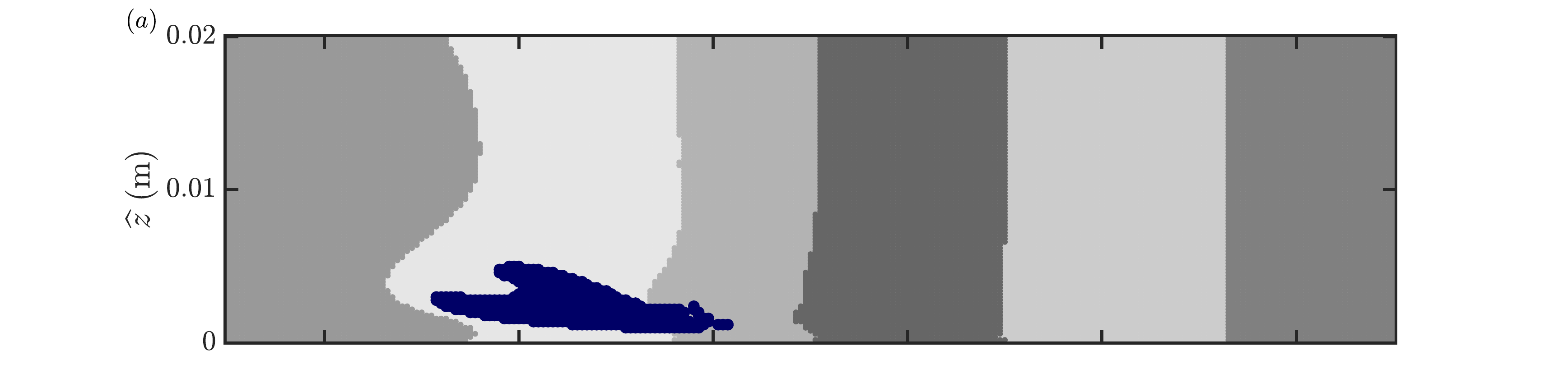}}
  \vspace{-0.cm}
  \centerline{\includegraphics[width=1.15\textwidth]{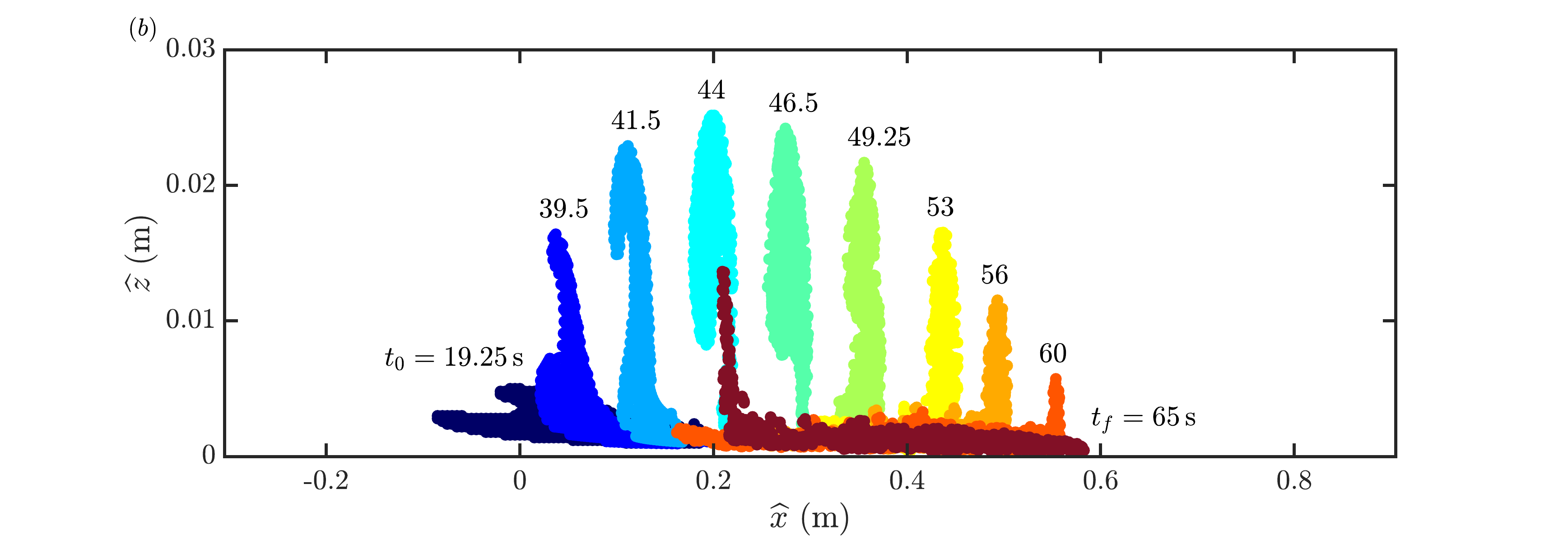}}
  \caption{\modif{$(a)$} Sample simulation clustering result for elements in the initial uniform grid.  Seven clusters have been identified, with the Lagrangian bolus cluster represented in \modif{dark blue}. $(b)$ The time evolution for the bolus cluster from $t_0=\SI{19.25}{s}$ to $t_f=\SI{65}{s}$, which is the entire bolus lifespan. (Video of the evolution of the figure is available in the Supplementary Material.)}
\label{fig:fig5}
\end{figure}

Figure~\ref{fig:fig5}$(a)$ presents the initial position of the partitions assigned to each one of the seven clusters identified for the sample simulation discussed in~\S\ref{sec:density_perturbation}.  
The \modif{dark blue} cluster is identified by the method as the objective bolus: the one cluster composed of tracers \modif{that eventually entrain in a propagating vortex} and move upslope. 
The bolus cluster also corresponds to the cluster of maximum displacement of the center of \modtwo{volume} in the $\xt$ direction.  
Figure~\ref{fig:fig5}$(b)$ presents how the bolus cluster propagates in time\modif{, and we note that tracers maintain their cluster membership throughout the duration of the time interval}. 
The bolus consists of tracers that are initially spread horizontally \modif{along the slope.}
As the wave arrives, those tracers get lifted, are trapped in the compact vortex, move along with the vortex for approximately $\SI{0.4}{m}$ up the slope, and eventually the front tracers stagnate and the trailing tracers start to recede down slope.
All the elements identified as part of the bolus end up trapped inside of the vortex in intermediate times.  
While not the case for this example, it is possible that no bolus is detected by the algorithm, indicating that the shoaling internal wave does not result in effective transport.

\subsection{Bolus transport properties} \label{sec:bolus_properties}

The objective quantification of the bolus makes it possible to measure bolus transport properties such as size, shape, position of center of volume and velocity.  
Important bolus properties are here defined, applied to the sample simulation, and will be used in the parametric studies in \S\ref{sec:Parametric}. 
Position and velocity of the Lagrangian bolus propagating up the slope are also compared to the unbroken portion of the shoaling internal wave.


Let $I_b$ be the set of tracers identified as part of the bolus, so that $|I_b|=n_b$ is the number of passive tracers identified as the bolus. 
The tracer positions $\{\mathbf{x}_{i}\}$ are known in the uniformly sampled time interval $[t_0,t_f]$.  
For each time instance $t$, averaging the position of the tracers gives the position of the bolus center of \modtwo{volume}
\begin{equation}
     \modtwo{\mathbf{x}_{CV}}(t) = \frac1{n_b} \textstyle \sum_{i\in I_b}\mathbf{x}_{i}(t),
\end{equation}
which we will use to quantify the bolus trajectory for $t\in [t_0,t_f]$.
The positions of the horizontally foremost and trailing tracers inside the bolus, $\mathbf{x}_{+}(t)$ and $\mathbf{x}_{-}(t)$ respectively, are also tracked.
These positions will correspond to different tracers as time evolves and overturning inside the bolus takes place.  These properties allow us to track the bolus center of \modtwo{volume} and its horizontal extent from a Lagrangian perspective.


To understand the relationship between the bolus and the shoaling wave that produces it, the wave trough and crest are also tracked.  
Because the unbroken components of the shoaling do not transport material, the displacement of isopycnals is tracked to locate the wave crest and trough.  
While the propagating wave amplitude is maximized at the \modif{center of the pycnocline}, other isopycnals throughout the water column are also deformed accordingly with these deformations vertically aligned.  
To track the unbroken shoaling wave, we track an isopycnal that is well above the breaking region, which makes it possible to track the shoaling wave propagation smoothly even after the bolus is generated.
The first propagating minimum and maximum of the isopycnal correspond to the leading wave trough and crest horizontal position.  Tracking the position of the isopycnal minimum and maximum provide a full description of the horizontal position of trough and crest as a function of time.

\begin{figure}
  \centerline{\includegraphics[width=1.15\textwidth]{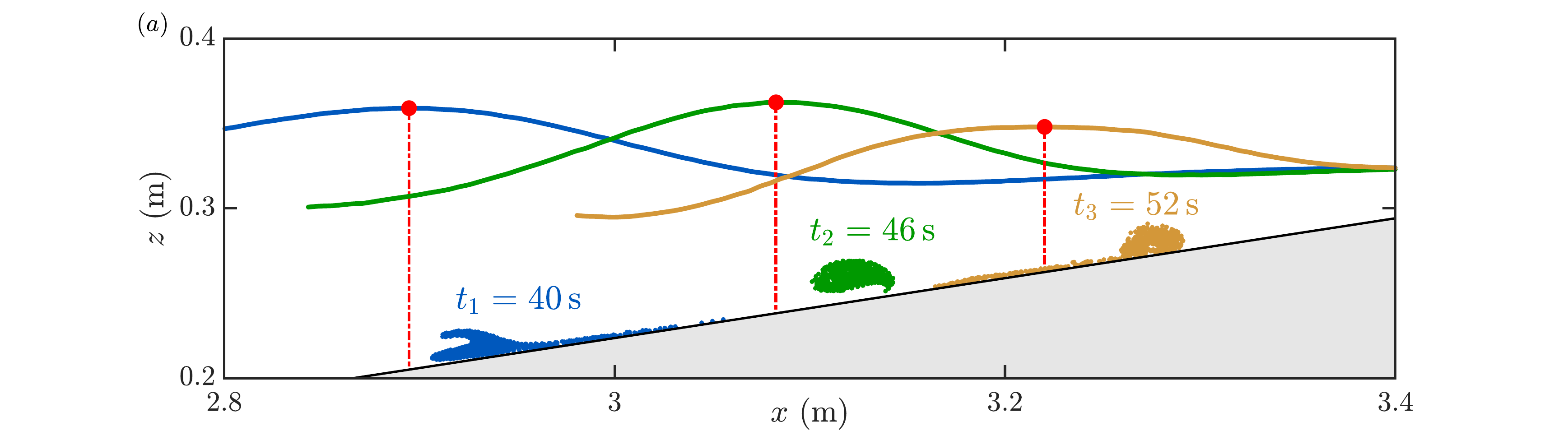}}
  \vspace{.2cm}
  \centerline{\includegraphics[width=1.15\textwidth]{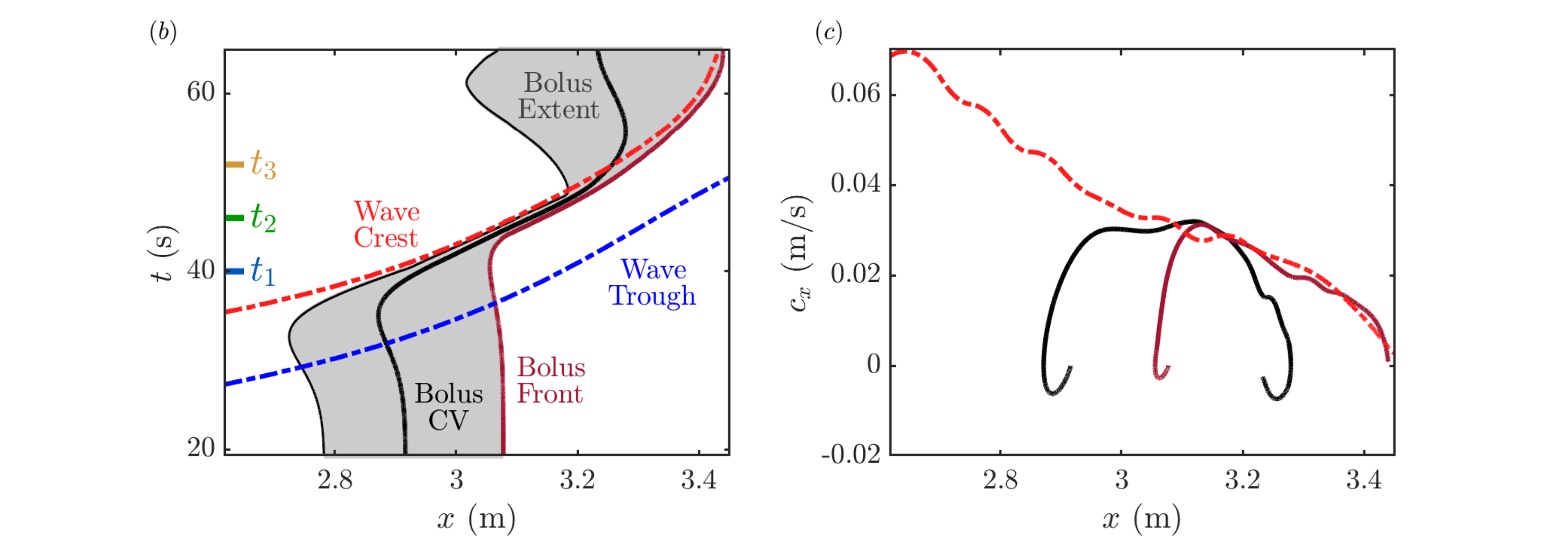}}
  \caption{\modif{ Kinematic comparison between the bolus and a co-propagating non-breaking isopycnal above the breaking region for $t\in[t_0,\,t_f]$. $(a)$ Bolus and amplified isopycnal for three time instances. The wave vertical positions and amplitudes are modified for illustration purposes.  $(b)$ The horizontal position as a function of time and  $(c)$ the horizontal speed as a function of horizontal position for the leading wave trough (dashed blue), the leading wave crest (dashed red), the bolus center of \modtwo{volume} (bold black), bolus trailing point (thin black) and foremost point (bold dark red). The isopycnal used to track the crest and trough in this plot was $\rho = \SI{1000.05}{kg/m^3}$, which has a neutrally buoyant height at $z=\SI{0.3635}{m}$. The tick marks in $(b)$ correspond to the times for which the bolus and isopycnals are plotted in $(a)$. }  }
  \label{fig:fig6}
\end{figure}


By tracking the extent of the bolus and the wave crest and trough, we can unveil how the location of the bolus relative to the wave evolves with time.  Bolus and wave kinematics are compared in figure~\ref{fig:fig6}.   \modif{A magnified version of the tracked isopycnal is presented in figure~\ref{fig:fig6}$(a)$ at three different time instances, together with the corresponding boluses, demonstrating that the front of the bolus is always between the wave crest and trough.} Figure~\ref{fig:fig6}$(b)$ presents the $(x,t)$ curves for the horizontal extent of the bolus and the shoaling wave.
For the bolus, the trajectories presented are those of the center of \modtwo{volume}, foremost and trailing point, with the horizontal extent of the bolus in gray.
For the leading wave, the trough and crest's trajectories are presented. 
As the wave trough approaches the slope, the trailing point of the bolus initially recedes while the front remains unaffected.  \modtwo{The wave crest then starts to approach, pushing the rear of the bolus forward and ultimately creates a vortex, which entrains all of the bolus tracers.
Finally, as depicted in figure~\ref{fig:fig6}$(a)$ for $t=\SI{52}{s}$, the bolus extent begins to increase, as some of the tracers are shed from the vortex, while the foremost tracers continue moving upslope with the decelerating wave crest.}
While the leading edge tracers remain ahead of the wave crest, the ejection of tracers from the vortex and their recession downslope causes the center of \modtwo{volume} to be left behind. 
Eventually the trailing points begin to move back up slope, at $t\approx\SI{62}{s}$, as the next shoaling wave arrives.

The horizontal speeds of the bolus center of \modtwo{volume}, foremost point and the leading crest are  presented in figure~\ref{fig:fig6}$(c)$. While the center of \modtwo{volume} moves slower than the crest, the correlated motion of the wave crest and the leading points is evident in the similar speed profiles with both undergoing the same deceleration as they move upslope where the water column narrows.  
Such a strong correlation between the Lagrangian bolus and the wave crest is not necessarily a result one would expect: the crest position is directly extracted from the density field, well above the breaking region, and the bolus is quantified by the movement of passive tracers identified as a coherent structure with the spectral clustering method. 
Nevertheless, the kinematics of both structures seem to consistently match, not only for the sample case presented, but for all parameter sweeps, which is an indicator that shoaling internal wave signatures close to the surface may possibly be used to estimate bolus location and speed. 
Given the consistency of the correlation between the crest and the bolus leading edge throughout the different cases analyzed, we will not demonstrate it for each parameter sweep.


Instead, for a direct comparison between simulations with different parameters, we quantify how far upslope material is transported by using the trajectory of the bolus center of \modtwo{volume} in the rotated reference frame, $\xt_{CV}(t)$.  The maximum displacement upslope relative to the initial position, $D_b$, is
\begin{equation}\label{eq:D_b}
    D_b = \max_{t \in [t_0,t_f]} \left\{ \xt_{CV}(t) - \xt_{CV}(t_0) \right\}.
\end{equation}
For the sample case in figure~\ref{fig:fig5}$(b)$, $D_b$ equals $\SI{0.402}{m}$ and is achieved at $t=\SI{56}{s}$.  One could, alternatively, define this distance based on the position of the bolus foremost point, and that definition would provide a similar displacement upslope ($\SI{0.363}{m}$ for the sample case).
We consider the center of \modtwo{volume} to provide a better description of the Lagrangian bolus dynamics as it accounts for the complete coherent structure describing material trapped in the bolus.


Another quantity of interest is the amount of material being transported, which we quantify by the approximate area of the bolus. The bolus size, $S_b$, is estimated using the number $n_b$ of tracers in the bolus cluster:
\begin{equation}\label{eq:S_b}
    S_b = n_b\,\Delta \xt_p\, \Delta \zt_p,
\end{equation}
where $\Delta \xt_p = \SI{5e-3}{m}$ and $\Delta \zt_p = \SI{2e-4}{m}$ are the initial distances between tracers. 
Because the number of tracers defining the bolus is time-invariant, the bolus size $S_b$ is not a function of time, as is required for the incompressible flow.  
In figure~\ref{fig:fig5}$(a)$, the \modif{dark blue} cluster representing the bolus has size  $\SI{5.14e-4}{m^2}$.  

The initial shape of the bolus at $t_0$ is important as it describes the portion of the fluid domain transported up the slope.
For the presented sample case, it is initially a thin sliver of fluid just above the boundary with an aspect ratio of approximately 60:1.  
However, there is a ``hook'' on top of the sliver, localized around $(0.05,0.003)\SI{}{m}$, which is also entrained into the vortex and carried upslope as presented in figure~\ref{fig:fig5}$(b)$.  
Both the size and shape of the bolus vary with system parameters, as will be first demonstrated in \S\ref{sec:effect_pycnocline}.

\modtwo{
To contrast the properties of the boluses studied here with previous studies, we will introduce some relevant dimensionless parameters. These parameters depend on incoming wave characteristics that are computed here using the mid-depth isopycnal $\rho=\rho_{H/2}$ at time $t_0$: the amplitude $a$, defined as the crest vertical displacement, the wavelength $\lambda$, distance between the two zero isopycnal displacement points surrounding the crest, the wavenumber $k=2\pi/\lambda$, and the instantaneous wave speed $c_x$, computed by tracking the wave crest. 
The typically investigated dimensionless numbers are the topographic slope $s$, the wave steepness $ka$, \modtwo{the internal} Iribarren number $Ir = s/\sqrt{a/\lambda}$, which compares the steepness of the slope and that of the incoming wave~\citep{boegman05,sutherland13}, the wave Reynolds number $Re_w=c_xka^2/\nu$, the wave Richardson number $Ri_w = k\delta/(ka)^2$~\citep{thorpe68, troy05} and the wave Froude number $Fr=\omega a/(g'H/4)^{1/2}$, with $g'=g\D\rho/\rho_{H/2}$~\citep{moore16}.  
}

\subsection{Comparison to three-dimensional analysis}
\label{sec:2d_3d}

While \cite{venayagamoorthy07} demonstrated that a three-dimensional simulation yielded similar results to its two-dimensional counterpart, we must verify that the coherent structure analysis is not subject to greater coherence due to unrealistic two-dimensional turbulence~\citep{arthur14}.  To do so, an equivalent three-dimensional simulation was performed with identical physical parameters.  To perform this simulation, it was necessary to decrease the grid spatial resolution along the shelf to $\Delta_x=\SI{e-3}{m}$.  The three-dimensional domain was $\SI{0.15}{m}$ across the shelf with a spatial resolution of $\Delta_y=\SI{1.5e-3}{m}$ in the $\widehat y$ direction, corresponding to 101 mesh layers and approximately 40 million mesh cells.

\modtwo{Periodic boundary conditions were applied to the lateral boundaries, and initial perturbations were incorporated to the density field in the three-dimensional simulation, to trigger lateral instabilities during the breaking. The perturbation method was inspired by~\citet{arthur14,arthur16}, and adapted to the continuously stratified system by adding the noise term $\zeta'R$, where $\zeta'=\SI{e-3}{m}$ and $R\in[-1,1]$ is a uniformly distributed random number, to the vertical coordinate $z$ in \eqref{eq:rho(z)} when initializing $\rho_0(z)$, for every mesh cell.}
For the coherent structure analysis, the in-plane tracer spacing was not changed, and the across shelf tracer spacing was $\Delta \widehat y_p=\SI{0.01}{m}$, totaling 15 layers of tracers. For numerical purposes, the size of the grid of tracers considered in the analysis was reduced so that the total number of trajectories was $n\approx\;$190,000.  The pairwise distance function between trajectories was adapted to account for the periodicity in the $\widehat y$ direction.

\begin{figure}
\centerline{\includegraphics[width=1.05\textwidth]{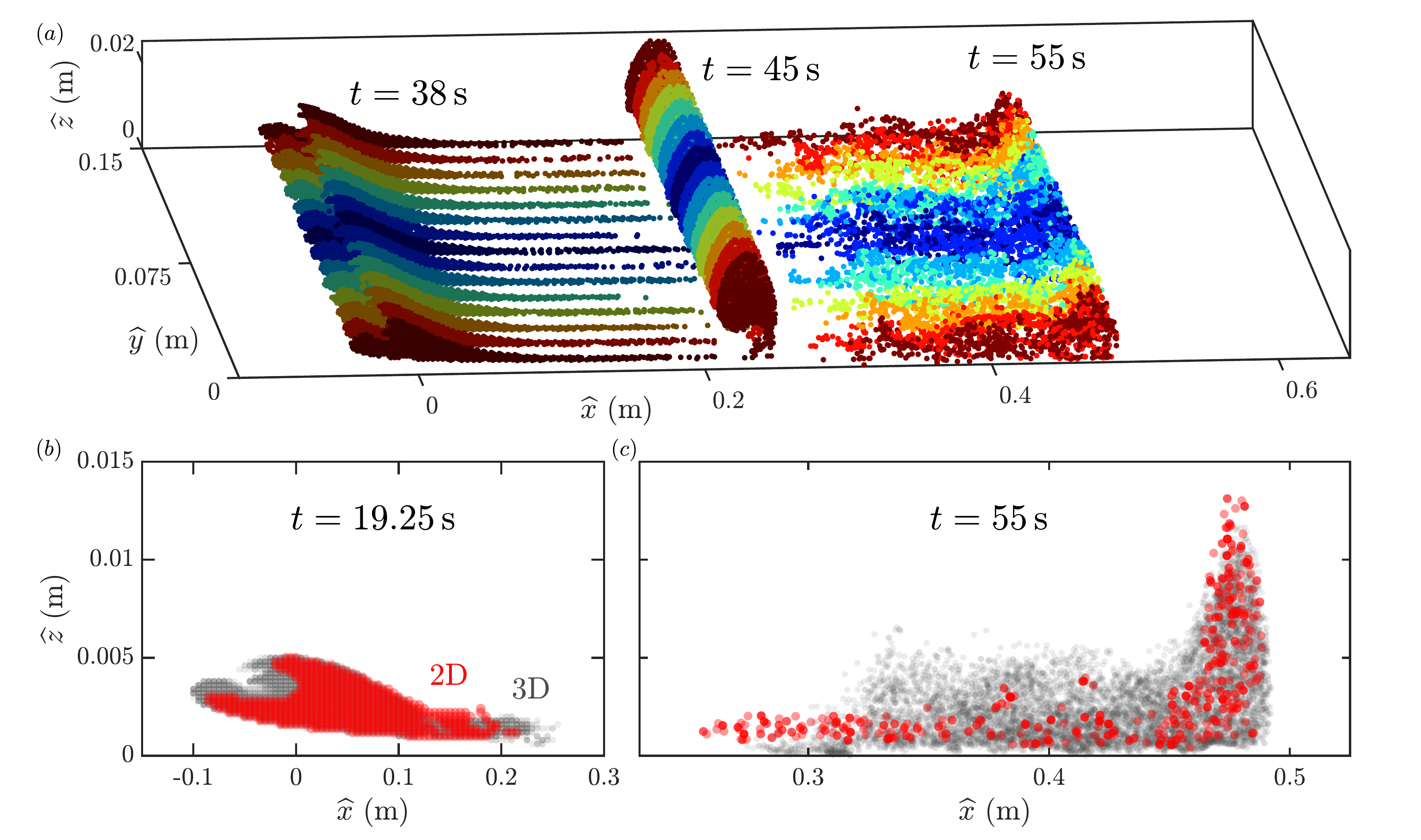}}
  \caption{\modtwo{$(a)$ Time evolution of the three-dimensional bolus. The tracer lateral coordinate $\widehat{y}$ at $t_0=\SI{19.25}{s}$ is indicated in color (from red to blue), and the different levels of shading indicate different time instances. $(b)$ Projected two-dimensional (red) and three-dimensional (gray) bolus tracers at the initial time $t_0$ and   $(c)$ at $t=\SI{55}{s}$. Transparency is used to indicate lower or higher concentration of tracers in the projected views.} }
\label{fig:fig7}
\end{figure}

The results of the three-dimensional coherent structure analysis are presented in figure~\ref{fig:fig7}.  
\modtwo{A perspective of the three-dimensional bolus is presented in~\ref{fig:fig7}$(a)$ for three time instances, $t=38$, $45$ and $\SI{55}{s}$. There is low lateral variability in the bolus shape for initial times, and only in later stages of the breaking the development of out-of-plane motion becomes more pronounced (see $t=\SI{55}{s}$).}
The initial position of the bolus tracers for the three-dimensional simulation align closely to the contour representing the two-dimensional cluster as demonstrated in figure~\ref{fig:fig7}$(b)$.  The three-dimensional bolus is slightly bigger, with an average cross-section area of \modtwo{$S_b=\SI{5.55e-4}{m^2}$} as compared to the $\SI{5.14e-4}{m^2}$ in the two-dimensional case.  In the three-dimensional case, the hook of the bolus becomes less pronounced \modtwo{and varies in shape for \modthree{different $\xt\zt$ cross-sections}}.  The propagation of these tracers over time allows the lateral variability to appear.  The bolus tracers at $t=\SI{55}{s}$ are presented in figure~\ref{fig:fig7}$(c)$.  While the front of the three-dimensional bolus is slightly further upslope, there is little difference in the distribution of the two- and three-dimensional tracers \modtwo{other than a wider distribution of particle positions off the sloping boundary in the three-dimensional case}.  The maximum displacement upslope is approximately the same in both cases,  \modtwo{$D_b=\SI{0.382}{m}$} for the three-dimensional case, compared to $\SI{0.402}{m}$ for the the two-dimensional case.

\modif{The similarities of the two- and three-dimensional results aligns well with the results of \cite{venayagamoorthy07}.  In their case, they use a background linear stratification which results in lower reduced gravity effects, similar to the broader pycnoclines in our study.  In their study, they identify lateral variation of the bolus primarily when the speed of the bolus slows to below the carrying wave speed.  \modtwo{As the wave shoals}, the water column height \modtwo{continues to decrease} and the carrying wave and bolus decelerate at similar rates as was observed in figure~\ref{fig:fig6}$(c)$.  \modtwo{This deceleration may be the reason why there is a delayed onset of the lateral instability, which produces an alternating pattern of lobes and clefts common to the head of gravity currents flowing over a no-slip boundary~\citep{simpson72}.}  \modtwo{While there is significant across-shelf variation at successive breaking events, lateral variability is minimal for the first bolus and therefore the two-dimensional model is physically realistic, as anticipated by \cite{aghsaee10}}.  This limits \modtwo{our} two-dimensional study to just the first bolus.  
}

\section{Bolus dependence on pycnocline thickness and other parameters} \label{sec:Parametric}

The coherent structure method provides an objective measurement of the bolus and can be used to understand the relationship between transport by boluses and system parameters.
While secondary system parameters such as the wave energy, the density change across the pycnocline and the topographic slope are varied one at a time, the effect of the pycnocline thickness, central to our investigation, is analyzed for all cases. 
Results demonstrating the importance and consequences of incorporating the pycnocline thickness $\delta$ to the model are presented in \S\ref{sec:effect_pycnocline}. The effects of secondary parameters on bolus formation and propagation are presented as follows: the incoming wave kinetic energy in \S\ref{sec:effect_energy}, the density change across the pycnocline in \S\ref{sec:effect_drho} and the topographic slope in \S\ref{sec:effect_slope}. \modif{The relationship between the bolus transport characteristics and relevant dimensionless parameters is investigated in \S\ref{sec:dimensionless_analysis}}. \modif{A summary of the simulations performed and the parameters used is presented in table~\ref{tab:tab1}.} \modtwo{For each simulation, the wave properties, the corresponding dimensionless parameters and further simulation details are provided in the Supplementary Material.}

\begin{table}
  \begin{center}
  \def~{\hphantom{1}}
  \begin{tabular}{c@{\hskip 10pt}c@{\hskip 10pt}c@{\hskip 10pt}c@{\hskip 10pt}c@{\hskip 10pt}c@{\hskip 10pt}c@{\hskip 10pt}}
      Section & Figure & Simulation \# & $\delta\,(\SI{}{m}$)   &  ${E_k}/{E_{k,0}}$  &  $\D\rho\,(\SI{}{kg/m^3}$)  & $s$\\[5pt]
      4.1 & 8 & 1 - 9 &  0.025 - 0.4   &   1   & 20 & 0.176\\
      4.2 & 10 & 10 - 27 &  0.025, 0.2, 0.4   &   1/8 - 4   & 20 & 0.176\\
      4.3 & 11 & 28 - 51 &  0.025 - 0.25   &   1   & 10 - 80 & 0.176\\
      4.4 & 12$(a$-$c)$ & 52 - 75 &  0.025, 0.2, 0.4   &   1   & 20 & 0.105, 0.176, 0.231\\
      4.4 & 12$(d$-$f)$ & 76 - 102 &  0.025 - 0.4   &   1   & 20 & 0.105 - 0.231
  \end{tabular}
  \caption{\modif{The bolus simulation cases considered in the respective sections and figures are presented here.  The pycnocline thickness $\delta$, the energy at the breaking point $E_k$, the change in density $\Delta \rho$, and the topographic slope $s$ are independently varied for each of the parameter studies. (An expanded version of this table is available in the Supplementary Material.)}}
  \label{tab:tab1}
  \end{center}
\end{table}

\subsection{Pycnocline thickness variation} \label{sec:effect_pycnocline}


While past internal wave bolus studies have focused on two-layer density stratifications, the impact of the pycnocline thickness $\delta$ determining how smoothly the density changes with depth is investigated here.  Nine different pycnocline thicknesses,
\begin{equation}
    \delta = 0.025,\,0.05,\,0.1,\,0.15,\,0.2,\,0.25,\,0.3,\,0.35,\,\SI{0.4}{m},
\end{equation}
are considered.  
Small $\delta$ values represent a thin, sharp transition of density and larger values represent broad, smooth transitions. 
The case $\delta = \SI{0.025}{m}$ is our approximate two-layer density system.  
A finer transition would require a higher resolution mesh and more computational time. 
The other extreme pycnocline thickness case, $\delta = \SI{0.4}{m}$, matches the height of the simulation domain. 
The case $\delta = \SI{0.2}{m}$, presented in \S\S\ref{sec:Numerical} and \ref{sec:Boluses}, is an intermediate value between these two extremes.

While the pycnocline thickness varies, the boluses are generated by equally energetic waves, with instantaneous kinetic energy at the breaking point $E_{k,0}$ (as discussed in \S\ref{sec:computational_approach}).  \modtwo{Because injecting a constant amount of energy may be simpler experimentally, that study is presented in the Supplementary Materials.}  $E_{k,0}$,  $\Delta \rho$ and $s$ are held constant for this study. \modtwo{As $\delta$ increases, the wave amplitude increases from $a=0.012$ to $\SI{0.016}{m}$, the wavelength remains approximately constant, $\lambda=0.68\pm \SI{0.01}{m},$ and the wave speed decreases from $c_x=0.112$ to $\SI{0.088}{m/s}$.   The wave steepness $ka$ varies from 0.11 to 0.15, the internal Iribarren number $Ir$  from 1.3 to 1.1, the wave Reynolds number $Re_w$ from 160 to 225, the Richardson number $Ri_w$ from 17 to 157 and the Froude number $Fr$ from 0.056 to 0.074, all of them monotonically with $\delta$.}

\begin{figure}
  \centerline{\includegraphics[width=1\textwidth]{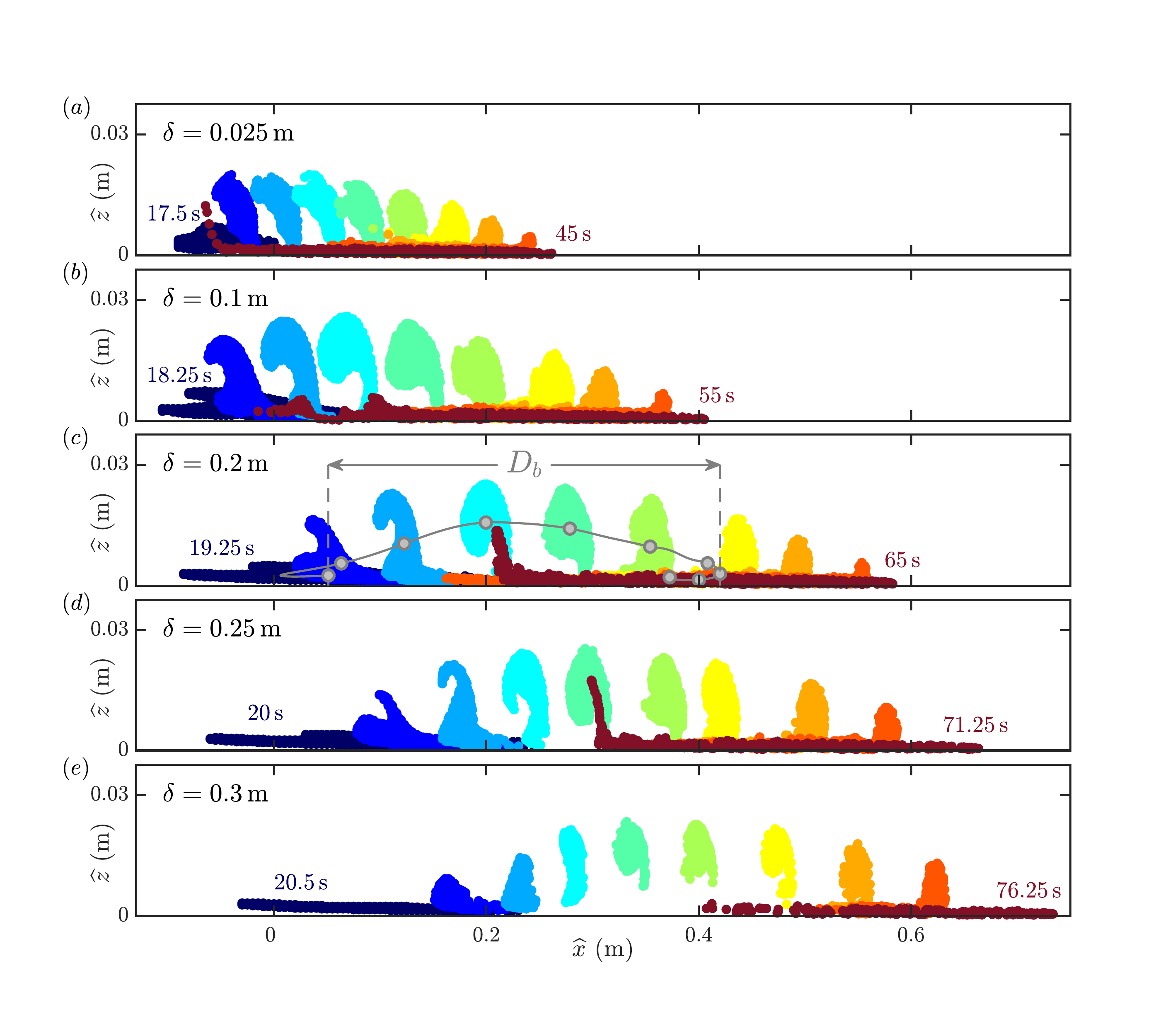}}
  \caption{Time evolution of the boluses from $t_0$ to $t_f$ for stratifications with pycnocline thicknesses $(a)$ $\delta=\SI{0.025}{m}$, $(b)$ $\SI{0.1}{m}$, $(c)$ $\SI{0.2}{m}$, $(d)$ $\SI{0.25}{m}$ and $(e)$ $\SI{0.3}{m}$. The trajectory of the bolus center of \modtwo{volume} for $\delta=\SI{0.2}{m}$ and the displacement upslope $D_b$ are illustrated in $(c)$. (Video of the evolution of the figure is available in the Supplementary Material.)}
\label{fig:fig8}
\end{figure}


The impact of the pycnocline thickness on the bolus dynamics is presented in figure~\ref{fig:fig8}. 
Snapshots of the bolus in the rotated frame $\Ot_{\xt\zt}$ are presented for selected times within the respective time span $[t_0,t_f]$.
For the sample case, in figure~\ref{fig:fig8}$(c)$, the trajectory of the center of \modtwo{volume} is presented in gray and the maximum displacement upslope $D_b$ defined in \eqref{eq:D_b} is illustrated.  
The values of $D_b$ and the bolus lifetime ($t_f-t_0$) both grow monotonically with $\delta$.  
A more complex relationship is observed between the bolus size $S_b$ and $\delta$: boluses are smaller for the approximate two-layer system, their sizes initially increase with $\delta$, reach a maximum, then shrink from this maximum as $\delta$ approaches the \modif{broadest pycnocline} case. 
It is also worth noting  that as $\delta$ increases the wave breaking happens later in time and the fluid trapped is  initially located further up the slope.

\begin{figure}
  \centerline{\includegraphics[width=1.05\textwidth]{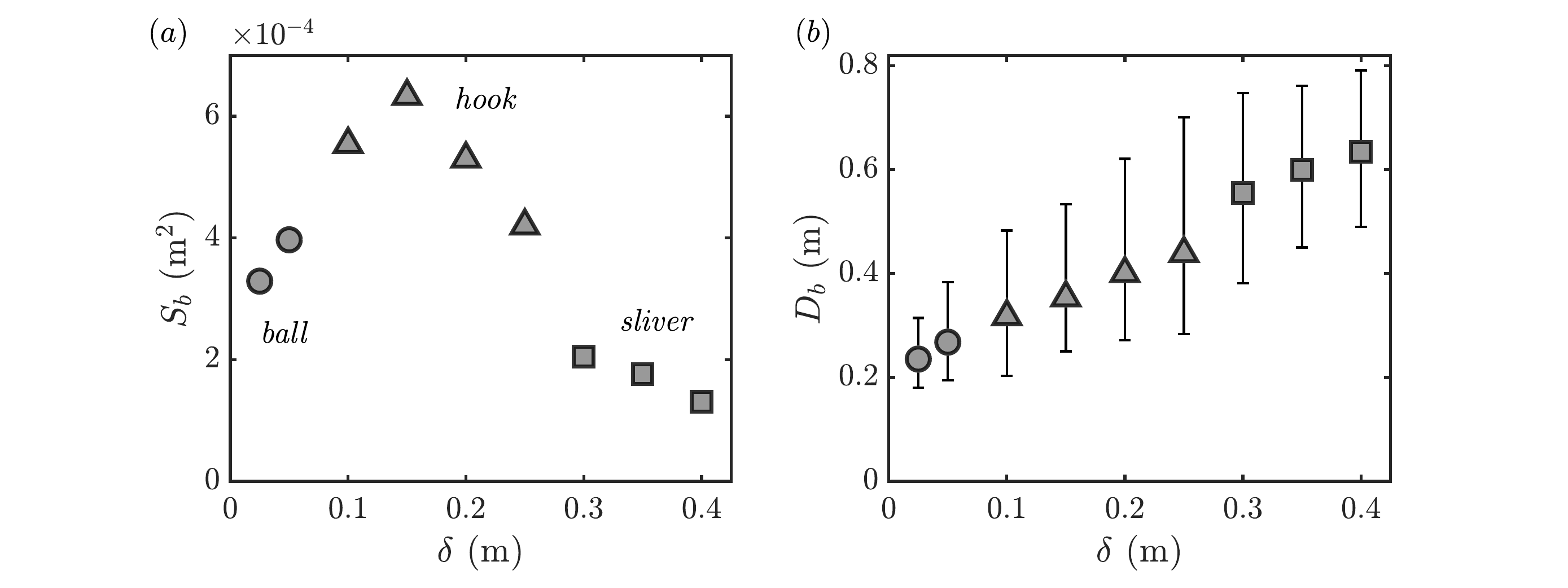}}
  \caption{$(a)$ Bolus size $S_b$ and $(b)$ maximum displacement upslope $D_b$ as a function of the pycnocline thickness, $\delta$, for mode-1 waves breaking with constant energy. \modif{The error bars represent the range of displacement within the bolus tracers.} Different markers are used to represent bolus shape categories: ball (circle), hook (triangle) and sliver (square).}
\label{fig:fig9}
\end{figure}


The bolus size $S_b$ and displacement upslope $D_b$ are presented in figure~\ref{fig:fig9}.  
The maximum $S_b$ is observed for $\delta=\SI{0.15}{m}$, while  $D_b$ monotonically increases with $\delta$.  
The largest bolus is approximately twice the size of the one observed for the approximate two-layer simulation and five times as large as the bolus produced in the \modif{broadest pycnocline} stratification.
The bolus obtained for the \modif{broadest pycnocline} case, however small, travels about three times as far as the two-layer bolus, with a lifetime more than twice as long.  \modtwo{Upper and lower bounds for the range of tracer displacement within the bolus are} also presented in figure~\ref{fig:fig9}$(b)$.  This range is smallest for the narrowest pycnoclines and gradually grows until $\delta=\SI{0.3}{m}$.  Both the upper and lower bound  increase monotonically.

\begin{figure}
  \centerline{\includegraphics[width=0.95\textwidth]{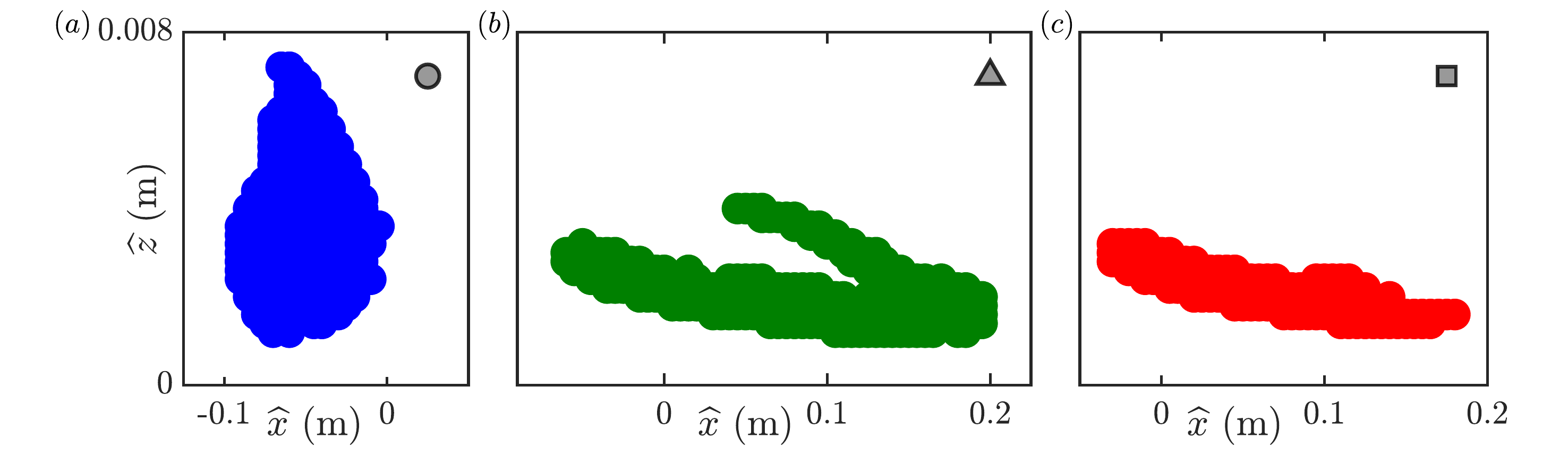}}
  \caption{$(a)$ \textit{Ball}, $(b)$ \textit{hook} and $(c)$ \textit{sliver} bolus shape categories represented by the bolus tracers position at time $t_0$. Shapes are obtained for $\delta=\SI{0.025}{m}$ (blue), $\SI{0.25}{m}$ (green) and $\SI{0.3}{m}$ (red). The aspect ratio is approximately 30:1. The markers in the top right of each plot are used to represent bolus shape categories.
  }
\label{fig:fig10}
\end{figure}

A jump in $S_b$ and $D_b$ occurs between boluses formed with~$\delta=\SI{0.25}{m}$ and~ $\delta=\SI{0.3}{m}$. This result relates to a difference in the shape of the bolus for each case. 
We refer to the bolus shape based on the initial geometry of the bolus tracers. 
In this paper, bolus shapes are divided into three categories: \textit{ball}, \textit{hook} and \textit{sliver}, which are illustrated in figure~\ref{fig:fig10}.
The \textit{ball} category, illustrated in figure~\ref{fig:fig10}$(a)$, is characterized by having a convex geometry, and is typical of thin pycnoclines, for which the breaking is more abrupt.
Increasing $\delta$ causes the initial bolus shape to take on the shape of a \quotes{hook}, which is eventually entrained in the vortex during the bolus propagation. 
The \textit{hook} category is well illustrated by the case $\delta = \SI{0.25}{m}$ in figure~\ref{fig:fig10}$(b)$, in which the hook is located around $(0.1,0.003)\,$m.
As $\delta$ grows further, only a horizontal stripe of tracers remain part of the bolus, and the hook is lost. The single horizontal stripe pattern defines the \textit{sliver} category, illustrated in figure~\ref{fig:fig10}$(c)$.  
The loss of the hook is a result of the propagating vortex not being strong enough to trap the material located in that area.  \modif{In fact, the maximum vorticity felt by the bolus monotonically decreases as $\delta$ increases.}
The transition from \textit{hook} to \textit{sliver} occurs somewhere between $\delta = \SI{0.25}{m}$ and~$\delta = \SI{0.3}{m}$.
This transition results in the bolus size decreasing by more than half.  \modif{This significant decrease in size may be the reason for the smaller range in displacements observed for $\delta = \SI{0.3}{m}$ compared to $\delta = \SI{0.25}{m}$.}
The corresponding increase in $D_b$ is related to the fact that the tracers that form the hook component are typically the first ones to be ejected from the vortex.

Drastic changes in the bolus size and displacement are found to be related to transitions in the bolus geometric properties described by these three categories. For this reason, for figure~\ref{fig:fig9} and all the $S_b$ and $D_b$ plots in the following sections, distinct bolus shapes are plotted using different symbols (as introduced in figure~\ref{fig:fig10}), so that transitions in shape are easily identifiable.

\subsection{Shoaling wave energy variation} \label{sec:effect_energy}


In the previous section, waves reached the slope with equal kinetic energy $E_{k,0}$.
Here, the influence of the shoaling wave energy on the bolus properties is investigated.  
\modtwo{We adjust the velocity forcing amplitude $a_f$  to produce waves with the energy scale factors}
\begin{equation}
    {E_k}/{E_{k,0}} = 1/8,\, 1/4,\, 1/2,\, 1,\, 2,\, 4.
\end{equation}
The upper bound for energy scale factors considered is set by computational limitations of direct simulations in highly energetic cases. 
The lower bound attempts to capture low-energy cases in which boluses do not form.  
To understand how the wave energy impacts the bolus propagation as a function of the pycnocline thickness, three thicknesses ($\delta=0.025,\,0.2$ and $\SI{0.4}{m}$) are considered for each energy scale factor.  For all simulations, $\Delta \rho$ and $s$ are held constant.  \modtwo{The wave amplitude increases with energy from $a=0.004$ to $\SI{0.033}{m}$, $\lambda$ remains approximately constant and $c_x$ is minimally impacted by the energy scale factor, being mostly determined by $\delta$.  Highly energetic waves are therefore steeper ($ka=0.04$ to $0.30$), resulting in $Ir$ varying from $2.19$ to $0.80$. $Re_w$ grows from 20 to 855, $Ri_w$ decreases with the energy and grows with $\delta$, spanning 6 to 1200, and $Fr$ grows with both $E_k$ and $\delta$ from 0.02 to 0.15.
}

\begin{figure}
  \centerline{\includegraphics[width=1.05\textwidth]{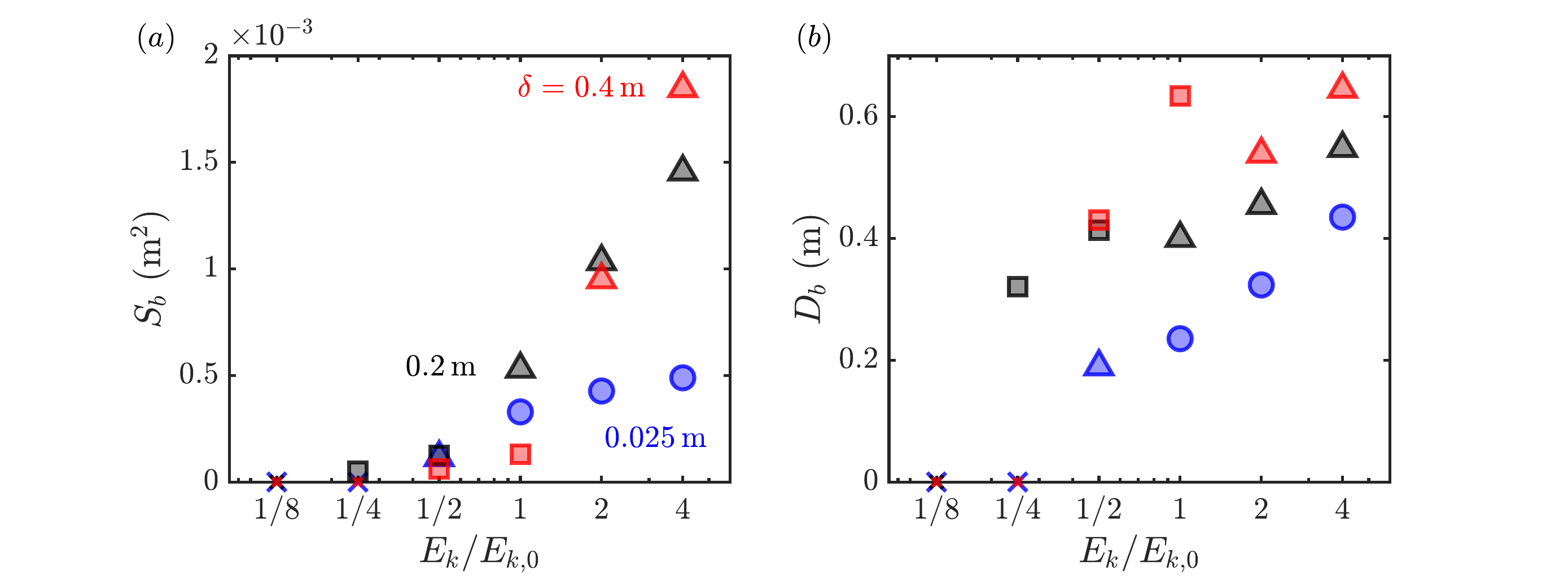}}
  \caption{$(a)$ Bolus size $S_b$ and $(b)$ maximum displacement upslope $D_b$ as a function of the energy factor for the shoaling wave. Simulations were performed with the pycnocline thicknesses $\delta=0.025$ (blue), $0.2$ (black) and $\SI{0.4}{m}$ (red). When no bolus is identified, \modif{a cross} is plotted at $S_b=\SI{0}{m^2}$ and $D_b=\SI{0}{m}$. Marker shapes are used to represent each bolus shape categories: ball (circle), hook (triangle) and sliver (square).
  }
\label{fig:fig11}
\end{figure}


Figure~\ref{fig:fig11} presents the relationship between the shoaling wave energy and the bolus properties.  
The bolus size dependence is presented in figure~\ref{fig:fig11}$(a)$ and the displacement dependence in figure~\ref{fig:fig11}$(b)$.
When no bolus is identified, the properties for that case are plotted in the graph as crosses with $S_b=\SI{0}{m^2}$, $D_b=\SI{0}{m}$. 
When the energy scale factor is $1/8$, none of the three pycnocline thicknesses produce a bolus.  
For ${E_k}/{E_{k,0}} = 1/4$, the only case producing a bolus is $\delta=\SI{0.2}{m}$.
For ${E_k}/{E_{k,0}} \geq 1/2$, all three cases of $\delta$ produce boluses, and $S_b$ increases with the wave energy.
The bolus size sensitivity to $E_k$ is weakest for the two-layer case, which is possibly related to the more abrupt breaking and quick mixing characteristic of this stratification.  The pycnocline thickness producing the largest and smallest bolus depends on $E_k$.

Varying the wave energy also results in transitioning between bolus shape categories, and transitions happen between different energy factors for different pycnocline thicknesses.
In particular, for $\delta=\SI{0.2}{m}$ a transition from sliver to hook-shaped boluses occurs when the energy factor is increased from $1/2$ to $1$, while this same transition for $\delta=\SI{0.4}{m}$ is observed between the energy factors $1$ to $2$.
This result is related to the fact that at higher energy the vortex has a greater circulation and is more capable of retaining particles that are moving with the bolus. 
The jumps in $S_b$ and $D_b$ resulting from these shape transitions produce counter-intuitive results: for example, the bolus displacement for $\delta=\SI{0.2}{m}$ is larger when the incoming wave has half of the energy compared to the baseline energy.  
The energy factor $1/2$ produces a sliver bolus that is approximately half the size of the hook bolus of energy ${E_{k,0}}$, so this bolus consists of fewer tracers that travel a longer distance.  
Such hook-sliver transitions explain why there are two clearly different growth rates observed for $S_b$ and $D_b$ for $\delta=0.2$ and  $\SI{0.4}{m}$, as the smaller sliver-shaped boluses travel longer distances compared to the larger hook-shaped boluses.

\subsection{Density change variation} \label{sec:effect_drho}


The stratification was varied in the previous sections by changing the pycnocline thickness, while keeping the density change parameter $\D\rho$ constant.  
Variations in $\D\rho$ have not been investigated in previous bolus studies. However, when considering typical ocean stratification profiles, $\D\rho$ varies and its value may have an impact on the resulting  bolus dynamics.  
Here, we vary the stratification by imposing density changes
\begin{equation}
    \D\rho = 10,\,20,\,40,\,\SI{80}{kg/m^3}.
\end{equation}
The choice of not studying values greater than $\SI{80}{kg/m^3}$ is related to the limitations of the Boussinesq approximation when describing stronger density variations. 
For the lower bound,  smaller changes than $\SI{10}{kg/m^3}$, for the fixed value of $\omega$, would result in a stratification with buoyancy frequency $N(z)<\omega$ for all $z$, resulting in evanescent waves.  
To understand how the sensitivity to $\D\rho$ varies across different pycnocline thicknesses, each density change is studied for the values of $\delta$ from $0.025$ to $\SI{0.25}{m}$. In all cases, the energy of the shoaling wave is $E_{k,0}$ and the topographic slope $s$ is held constant.

\modtwo{The combination of different density changes and pycnocline thicknesses has a significant impact \modthree{on all of the} wave characteristic: $a$ varies from 0.023 to $\SI{0.004}{m}$, $\lambda$  from 0.65 to $\SI{1.21}{m}$, and $c_x$ from 0.08 to $\SI{0.25}{m/s}$. For higher values of $\D\rho$, the waves have smaller amplitude and larger wavelength and speed.
For this study, the wave steepness varies within $ka=0.04$ and $0.22$, so that $Ir=0.95$ to $3.06$. $Re_w$ decreases with $\D\rho$, from 378 to 22, $Ri_w$ increases from 7 to 1980, while $Fr$ decreases from 0.15 to 0.01.
}


\begin{figure}
  \centerline{\includegraphics[width=1.05\textwidth]{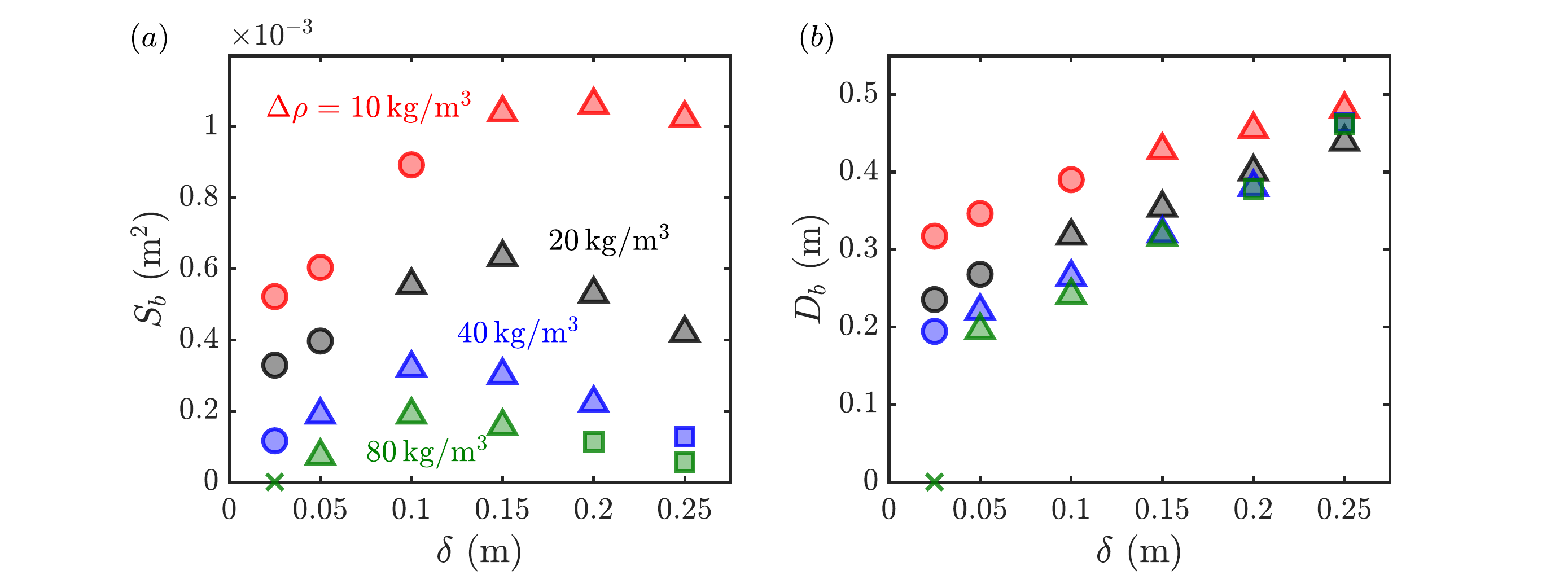}}
  \caption{$(a)$ Bolus size $S_b$ and $(b)$ maximum displacement upslope $D_b$ as a function of the pycnocline thickness, for variable density change, $\D\rho$. Simulations were performed with density changes $\D\rho=10$ (red), $20$ (black), $40$ (blue) and $\SI{80}{kg/m^{3}}$ (green). When no bolus is identified, \modif{a cross} is plotted at $S_b=\SI{0}{m^2}$ and $D_b=\SI{0}{m}$. Marker shapes are used to represent each bolus shape categories: ball (circle), hook (triangle) and sliver (square).
  }
\label{fig:fig12}
\end{figure}

The bolus size and displacement upslope are presented in figure~\ref{fig:fig12}.
Figure~\ref{fig:fig12}$(a)$ demonstrates that the pycnocline thickness that maximizes the bolus size depends on $\D\rho$. \modthree{The pycnocline thickness that results in the largest bolus increases from} $\delta=\SI{0.1}{m}$ for $\D\rho=\SI{80}{kg/m^3}$ to $\delta=\SI{0.2}{m}$ for $\D\rho=\SI{10}{kg/m^3}$.
For a strongly stratified system, the wave essentially reflects from the sloping wall, forming either a small bolus or no bolus, as is the case for $\D\rho=\SI{80}{kg/m^3}$ and $\delta=\SI{0.025}{m}$.  This case corresponds to one the greatest internal Iribarren numbers observed, $Ir=2.97$.
Figure~\ref{fig:fig12}$(b)$ confirms the trend previously observed in figure~\ref{fig:fig9}$(b)$: the displacement $D_b$ increases monotonically with $\delta$ for every $\D\rho$ value considered. 



One of the factors impacting $S_b$ for different stratifications is the speed of the internal wave, which is comparable to the speed of the bolus upslope (see figure~\ref{fig:fig6}).
Based on our simulations, slowly propagating boluses tend to trap more material inside the vortex than faster boluses. 
Another factor that plays a role in how much material is transported is the dynamics of the breaking itself.  Larger buoyancy frequencies result in the bolus edge destabilizing and mixing more quickly, resulting in less effective transport.

\subsection{Topographic slope variation} \label{sec:effect_slope}


Finally, to investigate the influence of the underlying topography, the topographic slope $s$, held at $0.176$ for the previous studies, is varied.
Keeping a constant density change $\D\rho = \SI{20}{kg/m^3}$ and shoaling wave energy $E_{k,0}$, the topographic slopes considered are
\begin{equation}
    s = 0.105,\,0.123,\,0.141,\,0.158,\,0.176,\,0.194,\,0.213,\,0.231,
\end{equation}
which correspond to slope angles of $6\degree$ to $13\degree$.
To ensure that the wave travels the same distance $L$ before breaking onslope, the domain geometry is adjusted so that the horizontal plane $z=H/2$ intersects the topography at $x=L$, as presented in figure~\ref{fig:fig2}.
Such a geometric constraint increases the domain size (and thus the number of mesh cells) for smaller slopes. The tested slopes are also constrained so that the topography does not intersect the domain inlet. For these two reasons $s$ is not \modthree{decreased} below $0.105$.  
To identify the relationship between bolus transport, topographic slope and pycnocline thickness, the thicknesses $\delta=0.025,\,0.2$ and $\SI{0.4}{m}$ are tested with each of the slopes.  
Three slopes corresponding to the extremes and the baseline case ($s=0.105,\,0.176$~and~$0.231$) are also tested with all nine pycnocline thicknesses.
\modtwo{While varying the topographic slope $s$ does not alter the properties of the incoming wave nor the dimensionless parameters $Re_w$, $Ri_w$ or $Fr$ from \S\ref{sec:effect_pycnocline}, changes in $s$ affect the internal Irribaren number $Ir$, which spans the range 0.67 through 1.71 as $s$ is increased.}


\begin{figure}
  \centerline{\includegraphics[width=1.17\textwidth]{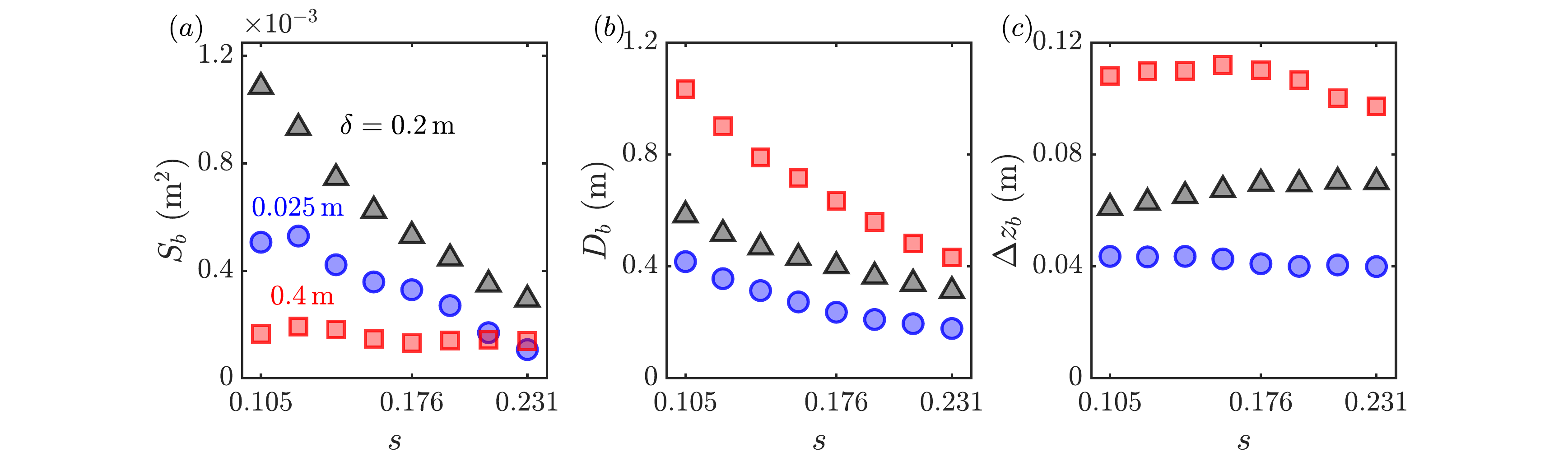}}
  \centerline{\includegraphics[width=1.17\textwidth]{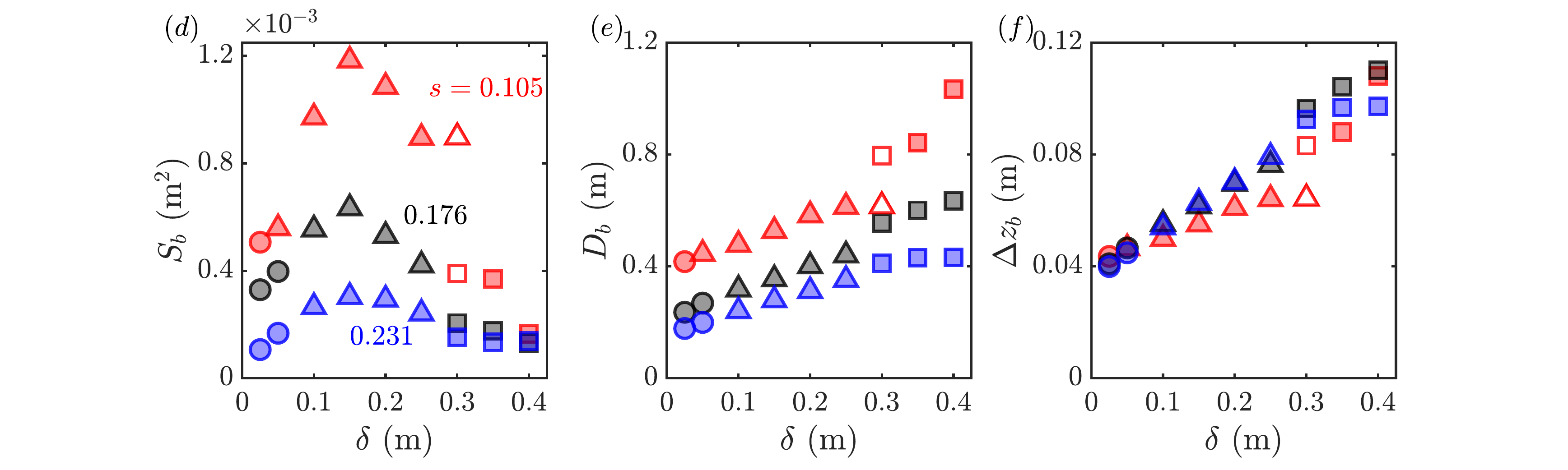}}
  \caption{ 
  \modif{$(a)$ Bolus size $S_b$, $(b)$ maximum displacement upslope $D_b$ and $(c)$ maximum vertical displacement $\D z_b$ as a function of the topographic slope $s$ for the pycnocline thicknesses $\delta=0.025$ (blue), $0.2$ (black), $\SI{0.4}{m}$ (red).
  $(d)$ Bolus size $S_b$, $(e)$ maximum displacement upslope $D_b$ and $(f)$ maximum vertical displacement $\D z_b$ as a function of the pycnocline thickness $\delta$, for slopes $s=0.105$ (red), $0.176$ (black), $0.231$ (blue).  Marker shapes are used to represent each bolus shape categories: ball (circle), hook (triangle) and sliver (square).
  Empty red markers for $s=0.105$, $\delta=\SI{0.3}{m}$ represent two possible values for the bolus at the hook-sliver transition.}}
  
\label{fig:fig13}
\end{figure}

The slope variation results are presented in figure~\ref{fig:fig13}. Initially, we examine the direct dependence on the slope $s$ for the three values of $\delta$ (figure~\ref{fig:fig13}$a$-$c$). 
The dependence of the bolus size $S_b$ on $s$ is presented in figure~\ref{fig:fig13}$(a)$. 
For $\delta=0.025$ and $\SI{0.4}{m}$, maxima in bolus size are obtained at approximately $s=0.123$.  
For $\delta=\SI{0.2}{m}$, there is a monotonic decrease in the bolus size as the topographic slope is increased. 
Based on the behavior observed for the extreme cases, there must be a slope shallower than $s=0.105$ that maximizes the size of the bolus for the case $\delta=\SI{0.2}{m}$ because there is ultimately no bolus generated when there is no slope.
With respect to the displacement upslope $D_b$, presented in figure~\ref{fig:fig13}$(b)$, steeper slopes result in boluses whose displacement along the slope decreases monotonically with $s$. 
This same trend is not observed if, instead of the displacement upslope $D_b$, the quantity plotted is the effective vertical displacement $\D z_b$. 
While $D_b$ decreases with $s$, $\D z_b$ is observed to remain nearly constant as $s$ is varied.  
This nearly constant $\D z_b$ behavior indicates that the stratification alone plays a major role in setting $\D z_b$, and $D_b$ is the projection of $\D z_b$ along the slope, resulting in longer displacements for shallower slopes and weaker stratifications.


Finally, the direct dependence of $S_b$, $D_b$ and $\D z_b$ on $\delta$ \modthree{for three values} of the slope $s$ is presented in figure~\ref{fig:fig13}($d$-$f$).  
The bolus size dependence in figure~\ref{fig:fig13}$(d)$ demonstrates that for all three values of $s$, the maximum $S_b$ is obtained for $\delta=\SI{0.15}{m}$.  
That means that the stratification corresponding to maximum bolus size is independent of the topographic slope for the range of values considered.  
The displacement upslope is presented in figure~\ref{fig:fig13}$(e)$ and reveals that the displacement increases nearly monotonically  with $\delta$. Figure~\ref{fig:fig13}$(f)$ presents the corresponding vertical displacement $\D z_b$, and a close correspondence between the different $s$ cases is observed. 

It is important to notice that the topographic slope also plays a role in determining when the ball-hook or hook-sliver bolus shape category transitions occur with respect to the pycnocline thicknesses.
Also, note that the combination $s=0.105$ and $\delta=\SI{0.3}{m}$ leads to the single instance in this paper where the bolus is very close to the transitioning point between hook and sliver categories, and both outputs are plotted in figure~\ref{fig:fig13}$(d$-$f)$ with empty symbols.
The output from the method as described in \S\ref{sec:lagrangian} corresponds to the hook case, but when we verify the robustness of the analysis by adding an extra cluster in the \textit{K}-means step, the analysis produces a sliver-shaped bolus.  This was the only case where this robustness analysis modified the bolus.

\subsection{\modif{Dimensionless analysis}} \label{sec:dimensionless_analysis}

The parametric studies presented in \S\S\ref{sec:effect_pycnocline}-\ref{sec:effect_slope} focus on varying one physical parameter at a time and evaluating how that parameter, in conjunction with the pycnocline thickness, impacts the properties of the bolus. Previous efforts have utilized dimensionless parameters to provide arguments about how the bolus will break, properties of the bolus, and the amount of mixing induced by the breaking process~\modtwo{\citep{ venayagamoorthy07, aghsaee10, sutherland13, moore16, arthur17}.}  Most of these studies focused on shoaling internal solitary waves of depression in a two-layer or nearly-two-layer density stratification, where the forcing mechanism and the resulting wave properties can be directly controlled.  \modtwo{Because the mode-1 generated, nonlinear internal waves simulated here experience some dispersion as they propagate from the inlet, studying the isolated effects of the dimensionless numbers or the wave properties is particularly challenging.}  Nevertheless, it is possible to recast our physical studies in a dimensionless perspective to relate our findings to previous studies.


\begin{figure}
  \centerline{
  \includegraphics[width=.925\textwidth]{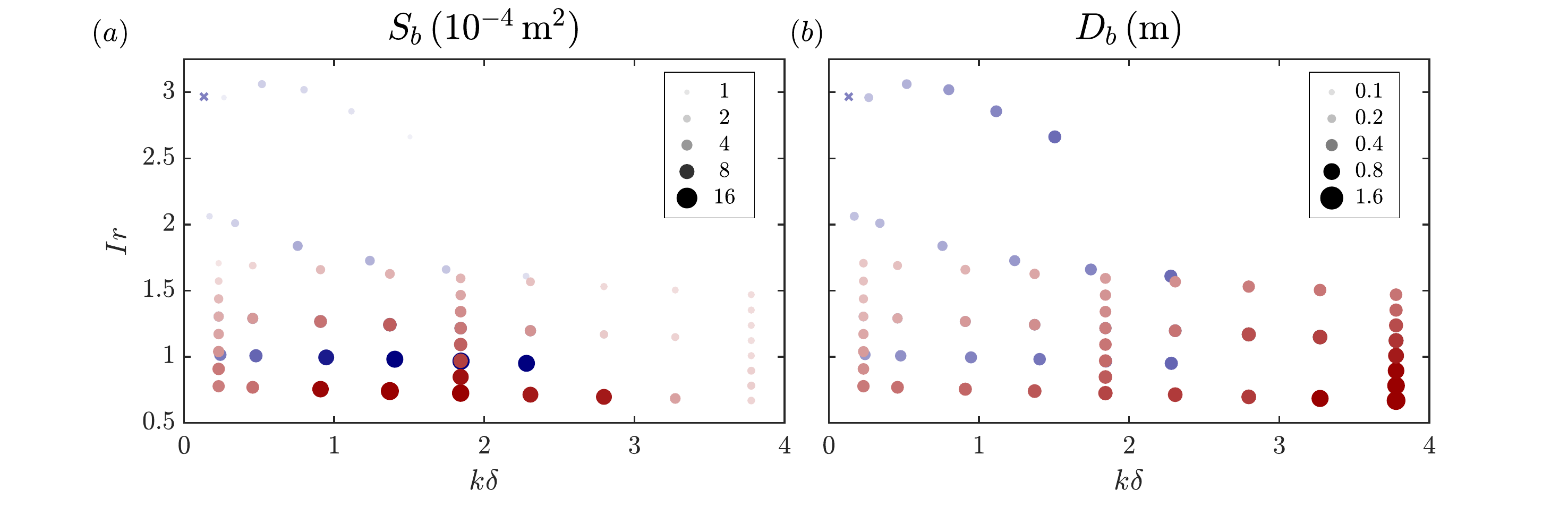}
  }
  \vspace{0.0cm}
  \centerline{
  \includegraphics[width=.925\textwidth]{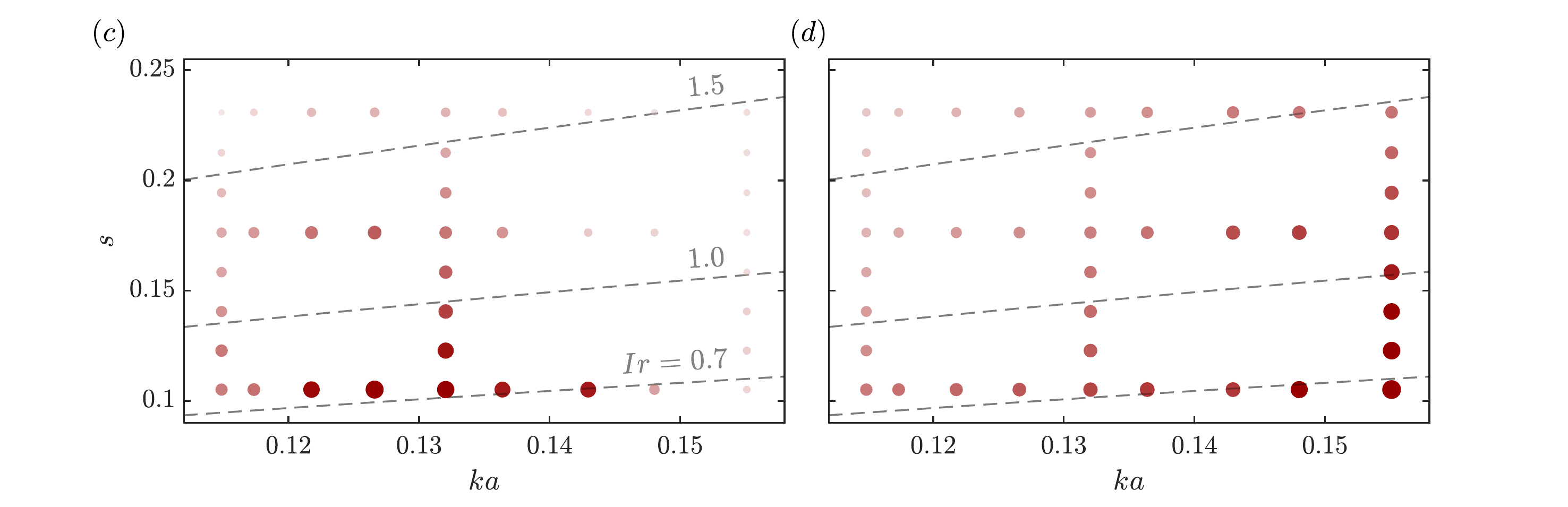}
  }
  \vspace{0.0cm}
  \centerline{  \includegraphics[width=.925\textwidth]{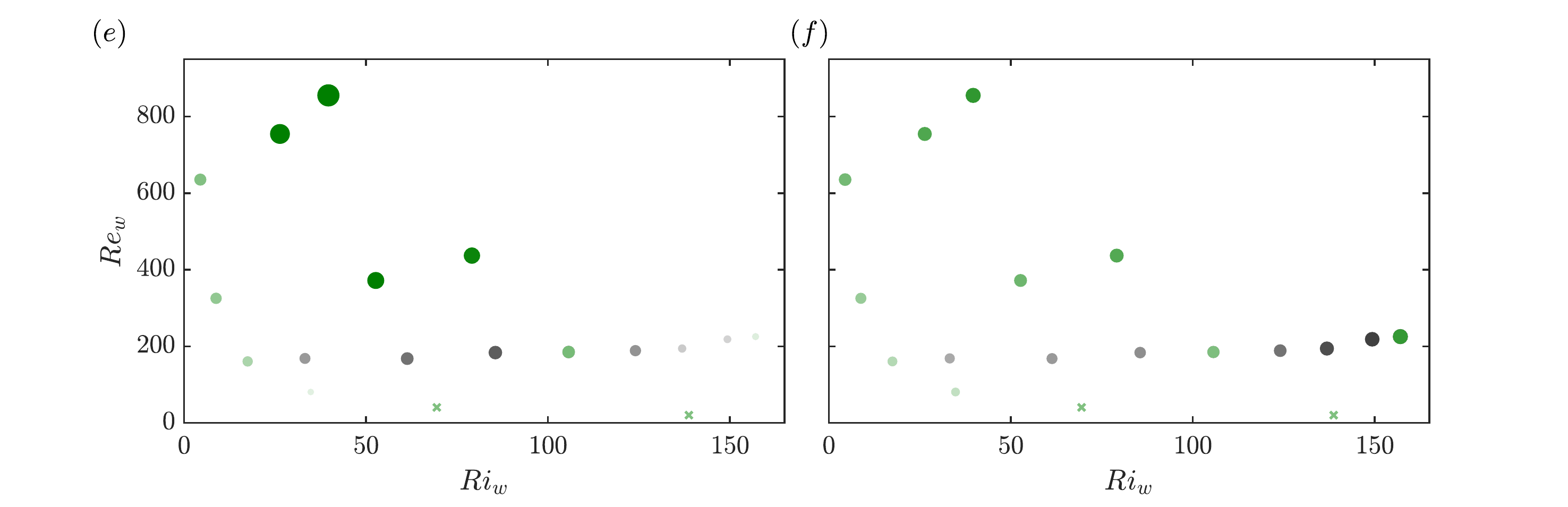}
  }
  \vspace{0.0cm}
  \centerline{
  \includegraphics[width=.925\textwidth]{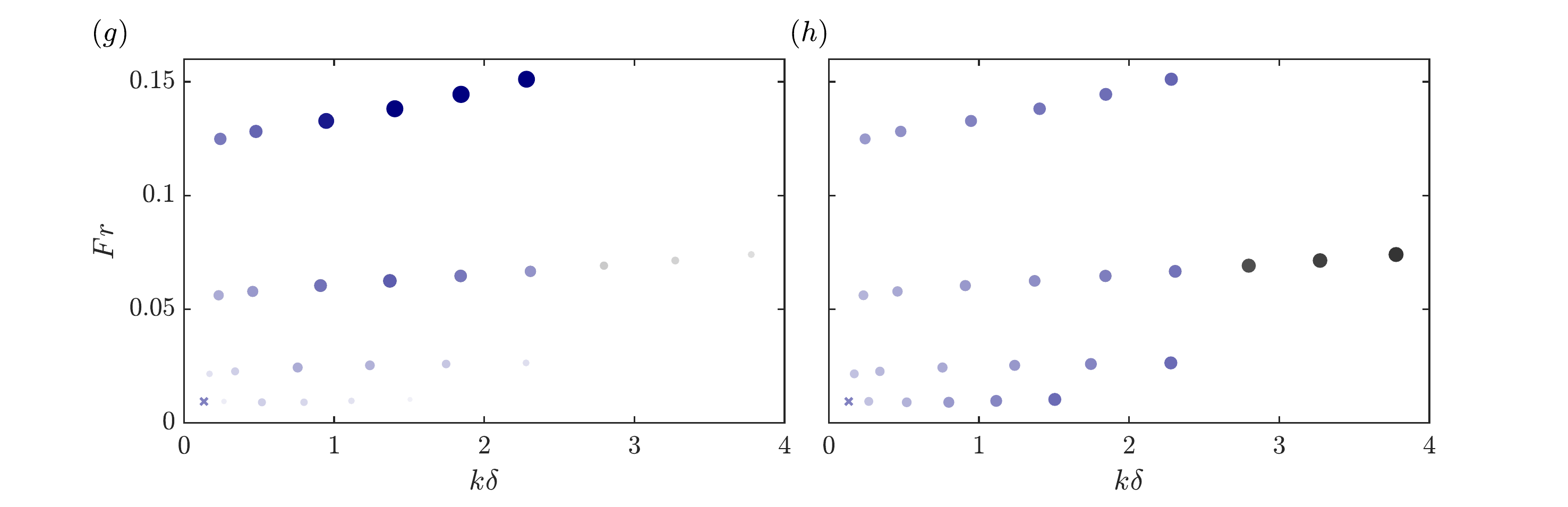}
  }
  \caption{\modif{Relationship between dimensionless parameters and bolus size $S_b$ and distance upslope $D_b$.  Marker sizes represent the magnitude of the plotted quantities, and colors represent different studies: pycnocline thickness variation (gray), energy variation (green), density change variation (blue), slope variation (red). Crosses represent cases for which no bolus is formed.}}
  \label{fig:fig14}
\end{figure}

The size and displacement dependence on pairs of dimensionless values defined in \S\ref{sec:bolus_properties} is presented in figure~\ref{fig:fig14}.  The first comparison  evaluates the relationship between $Ir$ and $k\delta$ in figure~\ref{fig:fig14}$(a$-$b)$.  Here, the size of the bolus decreases with increasing $Ir$ and there is an optimal $k\delta$ that produces the largest bolus.  The optimal $k\delta$ appears to decrease as $Ir$ increases.  The displacement of the bolus increases for increasing $k\delta$ and decreasing $Ir$.  This trend is observed in figure~\ref{fig:fig14}$(c$-$d)$ as $a$ is correlated to $\delta$.  This results in an optimal $ka$ for bolus size.  The displacement upslope increases with $ka$ and decreases as the topographic slope increases.  However, the trend in figure~\ref{fig:fig13}$(c)$ indicates that the vertical displacement of the bolus has a more complicated relationship with the slope.  Next, $Re_w$ and $Ri_w$ are compared in figure~\ref{fig:fig14}$(e$-$f)$.  These two numbers are most strongly varied by changing the energy at the breaking location $E_{k}$.  Based on the simulations performed, the size of the bolus increases with $Re_w$, but there is a non-monotonic trend with $Ri_w$.  The bolus displacement grows with both $Re_w$ and $Ri_w$.  It is interesting to note that for $Re_w<80$ no bolus is produced.  Finally, the dependence with $Fr$ is presented in figure~\ref{fig:fig14}$(g$-$h)$.  These figures demonstrate that the bolus size grows with $Fr$ but there is limited change in displacement.

The calculation of these dimensionless numbers allows us to relate the bolus breaking to those observed in other studies.  \cite{aghsaee10} numerically studied the breaking of two-layer waves for a range of topographic and wave slopes.  Our topographic slope ranges from $s=0.105$ to $0.231$ and \modtwo{the wave slope $a/\lambda$ ranges from $0.017$ to $0.025$}.  Based on their study, the corresponding wave breakers will be classified as ``surging'' events, which are characterized by minimal breaking of the incoming wave interface, surging of the wave upslope, and the production of a bolus.  This aligns well with what is observed in our simulations.
\modtwo{\citet{arthur16} also simulated the breaking of two-layer waves, observing surging breakers for their waves of lowest amplitude (case 1), which still have higher amplitude than any of the waves addressed here. Finally, \citet{arthur17} considered higher amplitude waves for different pycnocline thicknesses and classified all of their breakers as ``collapsing,'' which falls within the appropriate two-layer breaker classification, despite the pycnocline thickness variation. This matches our observation that changing the pycnocline thickness does not seem to modify the breaking classification.}

\section{Conclusions} \label{sec:Conclusions}



We have presented how the pycnocline thickness impacts internal wave bolus dynamics by applying Lagrangian coherent structure techniques to objectively identify and quantify the bolus transport properties.  
By modelling the continuous density stratification as a hyperbolic tangent profile with a tunable pycnocline thickness $\delta$, the dynamics of the bolus and the resulting transport property trends reveal significant differences when compared to two-layer density stratification models.
To identify the bolus in a continuous stratification, a spectral clustering method was used to detect the coherent structure underlying material transport.
The results obtained indicate that boluses tend to be larger and travel for longer distances in continuously stratified systems like those found in the ocean.
The parametric study reveals the relationship between transport properties of boluses and the pycnocline thickness, incoming wave energy, density change in the pycnocline and topographic slope.
Non-monotonic trends of the bolus size with these parameters indicate that there may be situations where the bolus size can be maximized.


For a laboratory-scale system with a continuous density stratification controlled by the pycnocline thickness $\delta$, the corresponding fundamental vertical mode internal wave was numerically forced to generate an internal wave that propagates towards a constant-slope topography.  
Direct simulations of the wave shoaling onto the topography produced the Eulerian velocity field modeling the bolus formation and propagation up the slope.
This velocity field was used to compute Lagrangian trajectories of passive tracers, which were processed by a spectral clustering method~\citep{hadjighasem16} to objectively identify the bolus.  
Position and velocity of the Lagrangian bolus propagating up the slope were compared to isopycnals located well above the breaking, and it was determined that the leading-edge of the bolus propagates immediately in front of the wave crest driving the bolus up the slope.
This objective detection of tracers composing the bolus enabled the measurement of transport-related quantities such as size and displacement upslope. 
These quantities were then used as a basis of comparison for the parametric study.


When considering shoaling waves of equal energy, the pycnocline thickness played an important role in determining the bolus size and displacement upslope.
A maximum bolus size was obtained for an intermediate thickness case, and not for the extreme cases of the \modif{broadest pycnocline} or the two-layer density stratifications.
In addition to changing transport properties, the pycnocline thickness also impacted the shape of the bolus.  \modif{As $\delta$ increases, the shape category transitions from ball to hook to sliver, with the hook producing the largest bolus and the sliver propagating the furthest.  This pattern was observed for all parameter studies with constant energy.}
Intuitively, as the internal wave energy is increased, the bolus size and displacement should also increase. While this is generally the case, our study also reveals transitions in the bolus shape that correspond to a non-monotonic trend. 
These transitions occur at different wave energy levels for different pycnocline thicknesses.
The method also successfully identified scenarios where no bolus is formed, for low energetic waves.
The impact of the density difference across the pycnocline was also investigated, and it was observed that boluses in stronger stratifications are larger for thinner pycnoclines, while boluses in weaker stratifications are larger for broader thicknesses.
A variation of the topographic slope concluded that, for the range of slopes \modif{studied}, shallower slopes allow boluses to propagate longer distances upslope. 
However, the vertical displacement of the bolus appears to be driven by the particular stratification and \modif{is} not dependent on the slope itself. 
For boluses formed on different slopes, the bolus size was maximal \modif{for the same pycnocline thickness case}, independent of the slope.  \modif{The simulation results were recast in dimensionless parameters revealing a non-monotonic relationship between $k\delta$ and the size of the bolus.  The displacement upslope increased with $k\delta$, $ka$, $Ri_w$, $Re_w$, and decreased with $s$ and $Ir$.  Finally, based on the dimensionless numbers, our boluses fall into the surging category based on \citet{aghsaee10}, which matches the qualitative behaviour.}


The results presented in this paper demonstrate that the incorporation of a more representative model for the density stratification that accounts for the pycnocline thickness will lead to different transport characteristics than models using either the two-layer density or broad pycnocline stratification models. 
The results show, in particular, that both of those simplified models could be underestimating the amount of material being transported by internal wave boluses on the continental slopes.
In~\S\ref{sec:effect_pycnocline}, for instance, it was observed that the optimal pycnocline thickness produces a bolus twice as large as the nearly two-layer stratification, and five times as large as the \modif{broadest pycnocline} stratification. 
Larger differences were obtained for the variable energy and variable density change cases.
There were cases for which boluses were not formed for the nearly two-layer or \modif{broadest pycnocline} stratification cases, but were formed for \modtwo{intermediate stratifications}.
While the study presented here is based on a laboratory-scale system, the trends established in this paper may help identify when and where boluses may be important in the ocean. For instance, high energy waves in a stratification with relatively small density change, shoaling in gradual sloping regions, may produce large boluses that propagate long distances depending on the pycnocline thickness.


Extending our results to ocean scale systems is a natural next step.
An ocean model will require the incorporation of a turbulence model, Coriolis forces, a background tidal flow and realistic, more irregular topography geometries to estimate the velocity field.
A suitable ocean model should also incorporate uncertainty, and the robustness of the spectral clustering method implemented to uncertainties is still an open question.
Also, accounting for the inertia of the tracers can have an impact on the final set of trajectories and affect our conclusion for the amount of transport induced by boluses.
Another extension of this work would be to study the shoaling of internal waves and bolus propagation on a bed of sediments. 
Because it corresponds to a high shear event, the bolus induces a large upwelling velocity from the topography at the front of the bolus, which could result in high amounts of resuspension and sediment transport. 
Accounting for a deformable boundary and resuspension, using a topography that consists of loosely-packed sediments that rearrange with the flow, could yield insight into the gradual transformation of the underlying topography. 
While internal wave induced resuspension and sediment transport has been observed in the ocean~\citep{hosegood04,quaresma07} and numerically reproduced~\modthree{\citep{stastna08,bourgault14}}, an estimate of the influence of boluses remains to be determined.

\section*{Acknowledgments}
We thank Harry L. Swinney for helpful discussions. 
The numerical simulations presented here were run at the Texas Advanced Computing Center (TACC).  We thank the anonymous referees for their insightful suggestions.
This work was initiated during an internship supported by the Eiffel Excellence Scholarship Program and the Paris-Saclay University International Master’s Scholarship Program.

\bibliography{references.bib}

\begin{thebibliography}{}

\bibitem[Abernathey and Haller, 2018]{abernathey18}
Abernathey, R. and Haller, G. (2018).
\newblock Transport by lagrangian vortices in the eastern pacific.
\newblock {\em J. Phys. Oceanogr.}, 48:667--685.

\bibitem[Aghsaee et~al., 2010]{aghsaee10}
Aghsaee, P., Boegman, L., and Lamb, K.~G. (2010).
\newblock Breaking of shoaling internal solitary waves.
\newblock {\em Journal of Fluid Mechanics}, 659:289--317.

\bibitem[Aikman~III, 1984]{aikman84}
Aikman~III, F. (1984).
\newblock Pycnocline development and its consequences in the middle atlantic
  bight.
\newblock {\em Journal of Geophysical Research: Oceans}, 89(C1):685--694.

\bibitem[Alford, 2003]{alford03}
Alford, M.~H. (2003).
\newblock Redistribution of energy available for ocean mixing by long-range
  propagation of internal waves.
\newblock {\em Nature}, 423:159--162.

\bibitem[Alford et~al., 2015]{alford15}
Alford, M.~H., Peacock, T., MacKinnon, J.~A., Nash, J.~D., Buijsman, M.~C.,
  Centurioni, L.~R., Chao, S.-Y., Chang, M.-H., Farmer, D.~M., Fringer, O.~B.,
  et~al. (2015).
\newblock The formation and fate of internal waves in the south china sea.
\newblock {\em Nature}, 521(7550):65.

\bibitem[Allshouse et~al., 2016]{allshouse16}
Allshouse, M.~R., Lee, F.~M., Morrison, P.~J., and Swinney, H.~L. (2016).
\newblock Internal wave pressure, velocity, and energy flux from density
  perturbations.
\newblock {\em Phys. Rev. Fluids}, 1:014301.

\bibitem[Allshouse and Peacock, 2015]{allshouse15}
Allshouse, M.~R. and Peacock, T. (2015).
\newblock Lagrangian based methods for coherent structure detection.
\newblock {\em Chaos: An Interdisciplinary Journal of Nonlinear Science},
  25(9):097617.

\bibitem[Allshouse and Thiffeault, 2012]{allshouse12}
Allshouse, M.~R. and Thiffeault, J.-L. (2012).
\newblock Detecting coherent structures using braids.
\newblock {\em Physica D: Nonlinear Phenomena}, 241(2):95--105.

\bibitem[Arthur and Fringer, 2014]{arthur14}
Arthur, R.~S. and Fringer, O.~B. (2014).
\newblock The dynamics of breaking internal solitary waves on slopes.
\newblock {\em Journal of Fluid Mechanics}, 761:360--398.

\bibitem[Arthur and Fringer, 2016]{arthur16}
Arthur, R.~S. and Fringer, O.~B. (2016).
\newblock Transport by breaking internal gravity waves on slopes.
\newblock {\em Journal of Fluid Mechanics}, 789:93--126.

\bibitem[Arthur et~al., 2017]{arthur17}
Arthur, R.~S., Koseff, J.~R., and Fringer, O.~B. (2017).
\newblock Local versus volume-integrated turbulence and mixing in breaking
  internal waves on slopes.
\newblock {\em Journal of Fluid Mechanics}, 815:169--198.

\bibitem[Benjamin, 1968]{benjamin68}
Benjamin, T.~B. (1968).
\newblock Gravity currents and related phenomena.
\newblock {\em Journal of Fluid Mechanics}, 31(2):209--248.

\bibitem[Boegman et~al., 2005]{boegman05}
Boegman, L., Ivey, G.~N., and Imberger, J. (2005).
\newblock The degeneration of internal waves in lakes with sloping topography.
\newblock {\em Limnology and oceanography}, 50(5):1620--1637.

\bibitem[Boegman and Stastna, 2019]{boegman19}
Boegman, L. and Stastna, M. (2019).
\newblock Sediment resuspension and transport by internal solitary waves.
\newblock {\em Annual Review of Fluid Mechanics}, 51:129--154.

\bibitem[Bourgault et~al., 2014]{bourgault14}
Bourgault, D., Morsilli, M., Richards, C., Neumeier, U., and Kelley, D.~E.
  (2014).
\newblock Sediment resuspension and nepheloid layers induced by long internal
  solitary waves shoaling orthogonally on uniform slopes.
\newblock {\em Continental Shelf Research}, 72:21--33.

\bibitem[Cacchione and Wunsch, 1974]{cacchione74}
Cacchione, D. and Wunsch, C. (1974).
\newblock Experimental study of internal waves over a slope.
\newblock {\em Journal of Fluid Mechanics}, 66(2):223--239.

\bibitem[Carter et~al., 2005]{carter05}
Carter, G.~S., Gregg, M.~C., and Lien, R.-C. (2005).
\newblock Internal waves, solitary-like waves, and mixing on the monterey bay
  shelf.
\newblock {\em Continental Shelf Research}, 25(12-13):1499--1520.

\bibitem[Dauxois et~al., 2004]{dauxois04}
Dauxois, T., Didier, A., and Falcon, E. (2004).
\newblock Observation of near-critical reflection of internal waves in a stably
  stratified fluid.
\newblock {\em Physics of Fluids}, 16(6):1936--1941.

\bibitem[Dettner et~al., 2013]{dettner13}
Dettner, A., Paoletti, M.~S., and Swinney, H.~L. (2013).
\newblock {Internal tide and boundary current generation by tidal flow over
  topography}.
\newblock {\em Phys. Fluids}, 25:1--13.

\bibitem[Duda et~al., 2004]{duda04}
Duda, T.~F., Lynch, J.~F., Irish, J.~D., Beardsley, R.~C., Ramp, S.~R., Chiu,
  C.-S., Tang, T.~Y., and Yang, Y.-J. (2004).
\newblock Internal tide and nonlinear internal wave behavior at the continental
  slope in the northern south china sea.
\newblock {\em IEEE Journal of Oceanic Engineering}, 29(4):1105--1130.

\bibitem[Fringer and Street, 2003]{fringer03}
Fringer, O.~B. and Street, R.~L. (2003).
\newblock The dynamics of breaking progressive interfacial waves.
\newblock {\em Journal of Fluid Mechanics}, 494:319--353.

\bibitem[Froyland and Junge, 2018]{froyland18}
Froyland, G. and Junge, O. (2018).
\newblock Robust fem-based extraction of finite-time coherent sets using
  scattered, sparse, and incomplete trajectories.
\newblock {\em SIAM Journal of Dynamical Systems}, 17(2):1891--1924.

\bibitem[Froyland and Padberg-Gehle, 2015]{froyland15}
Froyland, G. and Padberg-Gehle, K. (2015).
\newblock A rough-and-ready cluster-based approach for extracting finite-time
  coherent sets from sparse and incomplete trajectory data.
\newblock {\em Chaos: An Interdisciplinary Journal of Nonlinear Science},
  25(8):087406.

\bibitem[Froyland et~al., 2010]{froyland10}
Froyland, G., Santitissadeekorn, N., and Monahan, A. (2010).
\newblock Transport in time-dependent dynamical systems: Finite-time coherent
  sets.
\newblock {\em Chaos: An Interdisciplinary Journal of Nonlinear Science},
  20:043116.

\bibitem[Fructus et~al., 2009]{fructus09}
Fructus, D., Carr, M., Grue, J., Jensen, A., and Davies, P.~A. (2009).
\newblock Shear-induced breaking of large internal solitary waves.
\newblock {\em Journal of Fluid Mechanics}, 620:1--29.

\bibitem[Gerkema and Zimmerman, 2008]{gerkema08}
Gerkema, T. and Zimmerman, J. T.~F. (2008).
\newblock An introduction to internal waves.
\newblock {\em Lecture Notes, Royal NIOZ, Texel}, 207.

\bibitem[Hadjighasem et~al., 2016]{hadjighasem16}
Hadjighasem, A., Karrasch, D., Teramoto, H., and Haller, G. (2016).
\newblock Spectral-clustering approach to lagrangian vortex detection.
\newblock {\em Physical Review E}, 93(6):063107.

\bibitem[Haller, 2002]{haller02}
Haller, G. (2002).
\newblock Lagrangian coherent structures from approximate velocity data.
\newblock {\em Phys. Fluids A}, 14:1851--1861.

\bibitem[Haller and Beron-Vera, 2013]{haller13}
Haller, G. and Beron-Vera, F.~J. (2013).
\newblock Coherent lagrangian vortices: The black holes of turbulence.
\newblock {\em Journal of Fluid Mechanics}, 731.

\bibitem[Haller et~al., 2016]{haller16}
Haller, G., Hadjighasem, A., Farazmand, M., and Huhn, F. (2016).
\newblock Defining coherent vortices objectively from the vorticity.
\newblock {\em Journal of Fluid Mechanics}, 795:136--173.

\bibitem[Ham and Iaccarino, 2004]{ham04}
Ham, F. and Iaccarino, G. (2004).
\newblock {Energy conservation in collocated discretization schemes on
  unstructured meshes}.
\newblock In {\em Annual Research Briefs}, pages 3--14. Stanford University.

\bibitem[Helfrich, 1992]{helfrich92}
Helfrich, K.~R. (1992).
\newblock Internal solitary wave breaking and run-up on a uniform slope.
\newblock {\em Journal of Fluid Mechanics}, 243:133--154.

\bibitem[Helfrich and Melville, 1986]{helfrich86}
Helfrich, K.~R. and Melville, W.~K. (1986).
\newblock On long nonlinear internal waves over slope-shelf topography.
\newblock {\em Journal of Fluid Mechanics}, 167:285--308.

\bibitem[Helfrich and Melville, 2006]{helfrich06}
Helfrich, K.~R. and Melville, W.~K. (2006).
\newblock Long nonlinear internal waves.
\newblock {\em Annu. Rev. Fluid Mech.}, 38:395--425.

\bibitem[Hosegood et~al., 2004]{hosegood04}
Hosegood, P., Bonnin, J., and van Haren, H. (2004).
\newblock Solibore-induced sediment resuspension in the faeroe-shetland
  channel.
\newblock {\em Geophysical Research Letters}, 31(9).

\bibitem[Inall et~al., 2000]{inall00}
Inall, M.~E., Rippeth, T.~P., and Sherwin, T.~J. (2000).
\newblock Impact of nonlinear waves on the dissipation of internal tidal energy
  at a shelf break.
\newblock {\em Journal of Geophysical Research: Oceans}, 105(C4):8687--8705.

\bibitem[King et~al., 2009]{king09}
King, B., Zhang, H.~P., and Swinney, H.~L. (2009).
\newblock {Tidal flow over three-dimensional topography in a stratified fluid}.
\newblock {\em Phys. Fluids}, 21:116601.

\bibitem[Klymak and Moum, 2003]{klymak03}
Klymak, J.~M. and Moum, J.~N. (2003).
\newblock Internal solitary waves of elevation advancing on a shoaling shelf.
\newblock {\em Geophysical Research Letters}, 30(20).

\bibitem[Kunze, 2003]{kunze03}
Kunze, E. (2003).
\newblock A review of oceanic salt-fingering theory.
\newblock {\em Progress in Oceanography}, 56(3-4):399--417.

\bibitem[Lamb, 2003]{lamb03}
Lamb, K.~G. (2003).
\newblock Shoaling solitary internal waves: on a criterion for the formation of
  waves with trapped cores.
\newblock {\em Journal of Fluid Mechanics}, 478:81--100.

\bibitem[Lamb, 2014]{lamb14}
Lamb, K.~G. (2014).
\newblock Internal wave breaking and dissipation mechanisms on the continental
  slope/shelf.
\newblock {\em Annual Review of Fluid Mechanics}, 46:231--254.

\bibitem[Lee et~al., 2018]{lee18}
Lee, F.~M., Allshouse, M.~R., Swinney, H.~L., and Morrison, P.~J. (2018).
\newblock {Internal wave energy flux from density perturbations in nonlinear
  stratifications}.
\newblock {\em J. Fluid Mech.}, 856:898--920.

\bibitem[Lee et~al., 2014]{lee14}
Lee, F.~M., Paoletti, M.~S., Swinney, H.~L., and Morrison, P.~J. (2014).
\newblock {Experimental determination of radiated internal wave power without
  pressure field data}.
\newblock {\em Phys. Fluids}, 26:046606.

\bibitem[Legg and Adcroft, 2003]{legg03}
Legg, S. and Adcroft, A. (2003).
\newblock Internal wave breaking at concave and convex continental slopes.
\newblock {\em Journal of Physical Oceanography}, 33(11):2224--2246.

\bibitem[Liu et~al., 2001]{liu01}
Liu, Q., Jia, Y., Liu, P., Wang, Q., and Chu, P.~C. (2001).
\newblock Seasonal and intraseasonal thermocline variability in the central
  south china sea.
\newblock {\em Geophysical Research Letters}, 28(23):4467--4470.

\bibitem[Long, 1956]{long56}
Long, R.~R. (1956).
\newblock Solitary wwaves in the one- and two-fluid system.
\newblock {\em Tellus}, 8(4):460--471.

\bibitem[Long, 1965]{long65}
Long, R.~R. (1965).
\newblock On the boussinesq approximation and its role in the theory of
  internal waves.
\newblock {\em Tellus}, 17(1):46--52.

\bibitem[Maderich et~al., 2001]{maderich01}
Maderich, V.~S., Van~Heijst, G. J.~F., and Brandt, A. (2001).
\newblock Laboratory experiments on intrusive flows and internal waves in a
  pycnocline.
\newblock {\em Journal of Fluid Mechanics}, 432:285--311.

\bibitem[Mahesh et~al., 2004]{mahesh04}
Mahesh, K., Constantinescu, G., and Moin, P. (2004).
\newblock {A numerical method for large-eddy simulation in complex geometries.}
\newblock {\em J. Comput. Phys.}, 197:215--240.

\bibitem[Masunaga et~al., 2017]{masunaga17}
Masunaga, E., Arthur, R.~S., Fringer, O.~B., and Yamazaki, H. (2017).
\newblock Sediment resuspension and the generation of intermediate nepheloid
  layers by shoaling internal bores.
\newblock {\em Journal of Marine Systems}, 170:31--41.

\bibitem[Masunaga et~al., 2015]{masunaga15}
Masunaga, E., Homma, H., Yamazaki, H., Fringer, O.~B., Nagai, T., Kitade, Y.,
  and Okayasu, A. (2015).
\newblock Mixing and sediment resuspension associated with internal bores in a
  shallow bay.
\newblock {\em Continental Shelf Research}, 110:85--99.

\bibitem[Maxworthy et~al., 2002]{maxworthy02}
Maxworthy, T., Leilich, J. S. J.~E., Simpson, J.~E., and Meiburg, E.~H. (2002).
\newblock The propagation of a gravity current into a linearly stratified
  fluid.
\newblock {\em Journal of Fluid Mechanics}, 453:371--394.

\bibitem[Mercier et~al., 2010]{mercier10}
Mercier, M.~J., Martinand, D., Mathur, M., Gostiaux, L., Peacock, T., and
  Dauxois, T. (2010).
\newblock {New wave generation.}
\newblock {\em J. Fluid Mech.}, 657:308--334.

\bibitem[Michallet and Ivey, 1999]{michallet99}
Michallet, H. and Ivey, G.~N. (1999).
\newblock Experiments on mixing due to internal solitary waves breaking on
  uniform slopes.
\newblock {\em Journal of Geophysical Research: Oceans}, 104(C6):13467--13477.

\bibitem[Moore et~al., 2016]{moore16}
Moore, C.~D., Koseff, J.~R., and Hult, E.~L. (2016).
\newblock Characteristics of bolus formation and propagation from breaking
  internal waves on shelf slopes.
\newblock {\em Journal of Fluid Mechanics}, 791:260--283.

\bibitem[Moum et~al., 2003]{moum03}
Moum, J.~N., Farmer, D.~M., Smyth, W.~D., Armi, L., and Vagle, S. (2003).
\newblock Structure and generation of turbulence at interfaces strained by
  internal solitary waves propagating shoreward over the continental shelf.
\newblock {\em Journal of Physical Oceanography}, 33(10):2093--2112.

\bibitem[Moum et~al., 2007]{moum07}
Moum, J.~N., Klymak, J.~M., Nash, J.~D., Perlin, A., and Smyth, W.~D. (2007).
\newblock Energy transport by nonlinear internal waves.
\newblock {\em Journal of Physical Oceanography}, 37(7):1968--1988.

\bibitem[Munk and Wunsch, 1998]{munk98}
Munk, W. and Wunsch, C. (1998).
\newblock {Abyssal recipes II: Energetics of tidal and wind mixing}.
\newblock {\em Deep Sea Res., Part I}, 45:1977--2010.

\bibitem[Osborne and Burch, 1980]{osborne80}
Osborne, A.~R. and Burch, T.~L. (1980).
\newblock Internal solitons in the andaman sea.
\newblock {\em Science}, 208(4443):451--460.

\bibitem[Paoletti et~al., 2014]{paoletti14}
Paoletti, M.~S., Drake, M., and Swinney, H.~L. (2014).
\newblock {Internal tide generation in nonuniformly stratified deep oceans}.
\newblock {\em J. Geophys. Res}, 119:1943--1956.

\bibitem[Pedlosky, 2013]{pedlosky13}
Pedlosky, J. (2013).
\newblock {\em Geophysical fluid dynamics}.
\newblock Springer Science \& Business Media.

\bibitem[Pineda, 1991]{pineda91}
Pineda, J. (1991).
\newblock Predictable upwelling and the shoreward transport of planktonic
  larvae by internal tidal bores.
\newblock {\em Science}, 253(5019):548--549.

\bibitem[Pineda, 1994]{pineda94}
Pineda, J. (1994).
\newblock Internal tidal bores in the nearshore: Warm-water fronts, seaward
  gravity currents and the onshore transport of neustonic larvae.
\newblock {\em Journal of Marine Research}, 52(3):427--458.

\bibitem[Quaresma et~al., 2007]{quaresma07}
Quaresma, L.~S., Vitorino, J., Oliveira, A., and da~Silva, J. (2007).
\newblock Evidence of sediment resuspension by nonlinear internal waves on the
  western portuguese mid-shelf.
\newblock {\em Marine Geology}, 246(2-4):123--143.

\bibitem[Ray and Mitchum, 1996]{ray96}
Ray, R.~D. and Mitchum, G.~T. (1996).
\newblock Surface manifestation of internal tides generated near hawaii.
\newblock {\em Geophysical Research Letters}, 23(16):2101--2104.

\bibitem[Sandstrom and Elliott, 1984]{sandstrom84}
Sandstrom, H. and Elliott, J.~A. (1984).
\newblock Internal tide and solitons on the scotian shelf: A nutrient pump at
  work.
\newblock {\em Journal of Geophysical Research: Oceans}, 89(C4):6415--6426.

\bibitem[Sandstrom and Oakey, 1995]{sandstrom95}
Sandstrom, H. and Oakey, N.~S. (1995).
\newblock Dissipation in internal tides and solitary waves.
\newblock {\em Journal of Physical Oceanography}, 25(4):604--614.

\bibitem[Schlitzer, 2000]{schlitzer00}
Schlitzer, R. (2000).
\newblock Electronic atlas of woce hydrographic and tracer data now available.
\newblock {\em Eos, Transactions American Geophysical Union}, 81(5):45--45.

\bibitem[Serra et~al., 2017]{serra17}
Serra, M., Sathe, P., Beron-Vera, F., and Haller, G. (2017).
\newblock Uncovering the edge of the polar vortex.
\newblock {\em Journal of the Atmospheric Sciences}, 74(11):3871--3885.

\bibitem[Sigman et~al., 2004]{sigman04}
Sigman, D.~M., Jaccard, S.~L., and Haug, G.~H. (2004).
\newblock Polar ocean stratification in a cold climate.
\newblock {\em Nature}, 428(6978):59.

\bibitem[Simpson, 1972]{simpson72}
Simpson, J.~E. (1972).
\newblock Effects of the lower boundary on the head of a gravity current.
\newblock {\em Journal of Fluid Mechanics}, 53(4):759--768.

\bibitem[Stastna and Lamb, 2008]{stastna08}
Stastna, M. and Lamb, K.~G. (2008).
\newblock Sediment resuspension mechanisms associated with internal waves in
  coastal waters.
\newblock {\em Journal of Geophysical Research: Oceans}, 113(C10).

\bibitem[Susanto et~al., 2005]{susanto05}
Susanto, R., Mitnik, L., and Zheng, Q. (2005).
\newblock Ocean internal waves observed.
\newblock {\em Oceanography}, 18(4):80.

\bibitem[Sutherland et~al., 2013]{sutherland13}
Sutherland, B.~R., Barrett, K.~J., and Ivey, G.~N. (2013).
\newblock Shoaling internal solitary waves.
\newblock {\em Journal of Geophysical Research: Oceans}, 118(9):4111--4124.

\bibitem[Thorpe, 1968]{thorpe68}
Thorpe, S.~A. (1968).
\newblock On the shape of progressive internal waves.
\newblock {\em Philosophical Transactions of the Royal Society of London.
  Series A, Mathematical and Physical Sciences}, 263(1145):563--614.

\bibitem[Thorpe, 1971]{thorpe71}
Thorpe, S.~A. (1971).
\newblock Experiments on the instability of stratified shear flows: miscible
  fluids.
\newblock {\em Journal of Fluid Mechanics}, 46(2):299--319.

\bibitem[Troy and Koseff, 2005]{troy05}
Troy, C.~D. and Koseff, J.~R. (2005).
\newblock The instability and breaking of long internal waves.
\newblock {\em Journal of Fluid Mechanics}, 543:107--136.

\bibitem[Venayagamoorthy and Fringer, 2006]{venayagamoorthy06}
Venayagamoorthy, S.~K. and Fringer, O.~B. (2006).
\newblock Numerical simulations of the interaction of internal waves with a
  shelf break.
\newblock {\em Physics of Fluids}, 18(7):076603.

\bibitem[Venayagamoorthy and Fringer, 2007]{venayagamoorthy07}
Venayagamoorthy, S.~K. and Fringer, O.~B. (2007).
\newblock On the formation and propagation of nonlinear internal boluses across
  a shelf break.
\newblock {\em Journal of Fluid Mechanics}, 577:137--159.

\bibitem[von Luxburg, 2007]{luxburg07}
von Luxburg, U. (2007).
\newblock A tutorial on spectral clustering.
\newblock {\em Statistics and computing}, 17(4):395--416.

\bibitem[Walter et~al., 2012]{walter12}
Walter, R.~K., Woodson, C.~B., Arthur, R.~S., Fringer, O.~B., and Monismith,
  S.~G. (2012).
\newblock Nearshore internal bores and turbulent mixing in southern monterey
  bay.
\newblock {\em Journal of Geophysical Research: Oceans}, 117(C7).

\bibitem[Wang et~al., 2007]{wang07}
Wang, Y.-H., Dai, C.-F., and Chen, Y.-Y. (2007).
\newblock Physical and ecological processes of internal waves on an isolated
  reef ecosystem in the south china sea.
\newblock {\em Geophysical Research Letters}, 34(18).

\bibitem[White and Helfrich, 2008]{white08}
White, B.~L. and Helfrich, K.~R. (2008).
\newblock Gravity currents and internal waves in a stratified fluid.
\newblock {\em Journal of Fluid Mechanics}, 616:327--356.

\bibitem[Wunsch and Ferrari, 2004]{wunsch04}
Wunsch, C. and Ferrari, R. (2004).
\newblock {Vertical mixing, energy and the general circulation of the oceans}.
\newblock {\em Annu. Rev. Fluid Mech}, 36:281--314.

\bibitem[Zhang et~al., 2008]{zhang08}
Zhang, H.~P., King, B., and Swinney, H.~L. (2008).
\newblock Resonant generation of internal waves on a model continental slope.
\newblock {\em Physical review letters}, 100(24):244504.

\bibitem[Zhang and Swinney, 2014]{zhang14}
Zhang, L. and Swinney, H.~L. (2014).
\newblock {Virtual seafloor reduces internal wave generation by tidal flow}.
\newblock {\em Phys. Rev. Lett.}, 112:104502.

\end{thebibliography}

\appendix

\clearpage
\section{Fundamental vertical mode waves}
\label{appendix:fundamental_mode}

\thispagestyle{empty}

\vspace{1em}

Here we present how to determine the fundamental vertical mode profile of an internal wave of frequency $\omega$ for an arbitrary stratification profile $\rho_0(z)$. The mode-1 wave contains most of the internal wave energy and corresponds to the component of largest wavelength.
The modal analysis relies on a linearization of \eqref{eq:NS-1}-\eqref{eq:NS-4} around a static background state satisfying~$\rho = \rho_0(z)$ and~$\textrm{d}p_0/\textrm{d}z = -\rho_0(z) g$. The small perturbations in $\rho$ and $p$ associated with the internal wave motion are denoted by primed fields, such that~$\rho = \rho_0(z)+\rho'(x,z,t)$ and~$p = p_0(z)+p'(x,z,t)$.

Under these assumptions, the linearized system becomes
\begin{eqnarray}
\label{eq:appendix_cont}
\pd{u}{x} + \pd{w}{z} & = & 0, \\
\label{eq:appendix_dudt}
\pd{u}{t} & = & -\frac{1}{\rho_{00}} \pd{p'}{x}, \\
\label{eq:appendix_dwdt}
\pd{w}{t} & = & -\frac{1}{\rho_{00}} \pd{p'}{z} - \frac{\rho'}{\rho_{00}}g, \\
\label{eq:appendix_drhodt}
\pd{\rho'}{t} - \frac{\rho_{00} N^2}{g} w & = & 0, 
\end{eqnarray}
where
\begin{equation}\label{eq:N}
N(z) = \sqrt{ - \frac{g}{\rho_{00}} \frac{\textrm{d}\rho_0 }{\textrm{d}z} }
\end{equation}
is the buoyancy frequency.
Combining \eqref{eq:appendix_cont}-\eqref{eq:appendix_drhodt}, we end up with a single equation to be solved for the vertical velocity:
\begin{equation}\label{eq:w}
\left(\frac{\partial^2}{\partial t^2}\boldnabla^2 + N^2(z) \frac{\partial^2}{\partial x^2}\right) w(x,z,t) = 0.
\end{equation}
Once $w$ is determined, all remaining perturbation profiles are resolved~\citep{gerkema08}.





A vertical mode correspond to a solution of \eqref{eq:w} that has the form of a plane wave propagating in the $x$ direction, with temporal frequency $\omega$ and wavenumber $k_x$:
\begin{equation}\label{eq:w_form}
w(x,z,t) = \mathrm{Re}\left\{ W(z)\exp(i(k_x x-\omega t))\right\},
\end{equation}
with $i$ is the imaginary number and $\mathrm{Re}\left\{\cdot \right\}$ denotes the real part of the complex argument.
Substitution of \eqref{eq:w_form} in \eqref{eq:w} results in the ordinary differential equation for $W(z)$
\begin{equation}\label{eq:W}
\frac{\textrm{d}^2W}{\textrm{d}z^2}+k_x^2\left(\frac{N^2(z)}{\omega^2}-1\right)W = 0.
\end{equation}
For $z\in[0,H]$, both top and bottom being horizontal surfaces, no-flux boundary conditions lead to $W(0) = W(H) = 0$.
For a Sturm-Liouville equation as \eqref{eq:W}, each solution $W^{(n)}(z)$ corresponding to a possible $\left(k_{x}^{(n)}\right)^2$ is unique and orthogonal to all others. 
The general solution of \eqref{eq:w} consists of the superposition of solutions in the form \eqref{eq:w_form}, with the $W^{(n)}(z)$ and $k_{x}^{(n)}$ corresponding to each vertical mode.
The fundamental mode, which we want to find, is the mode associated to the smallest wavenumber $k_{x}^{(0)}$.


For a general stratification $\rho_0(z)$, and thus a general $N^2(z)$, \eqref{eq:W} cannot be solved analytically. 
We solve \eqref{eq:W} numerically, by using a finite difference approach. 
Let $\mathbf{w} = (\mathrm{w}_0,\mathrm{w}_1,\dots,\mathrm{w}_{M+1})^\intercal$ be the vector with the discretized values of $W(z)$ at the points $z_0,\ldots,z_{M+1}$ uniformly distributed in $[0,H]$. We impose the homogeneous boundary conditions $\mathrm{w}_0 = \mathrm{w}_{M+1} = 0$ and use centered finite differences to approximate the second derivatives in \eqref{eq:W}. The discretized version of \eqref{eq:W} becomes the linear system
\begin{equation}\label{eq:eigenproblem}
\mathbf{A}\mathbf{w}+k_x^2\mathbf{B}\mathbf{w} = \mathbf{0},
\end{equation}
where
$$
A_{ij} = 
\begin{cases}
     -2/{\D z}^2 & \textrm{for } i=j,\\
     1/{\D z}^2 & \textrm{for } i=j\pm 1,\\
     0 & \textrm{otherwise}.
\end{cases}
\quad
B_{ij} =
\begin{cases}
     N^2(z_i)/\omega^2 - 1, & \textrm{for } i=j,\\
     0 & \textrm{otherwise}.
\end{cases}
$$
with $1\leq i,j \leq M$. \eqref{eq:eigenproblem} is a generalized eigenvalue problem that can be solved numerically, with generalized eigenvalues $k_{x}^{(n)}$ and eigenvectors $\mathbf{w}^{(n)}$. 

In the event of all $B_{ii}<0$, which corresponds to a stratification such that $N(z)<\omega$ everywhere,  \eqref{eq:eigenproblem} has no positive eigenvectors and the solutions will correspond to an evanescent wave~\citep{gerkema08}. The mode $\mathbf{w}^{(0)}$ corresponding to the smallest wavenumber $k_{x}^{(0)}$, numerically computed from \eqref{eq:eigenproblem}, is a discretization for the continuum fundamental mode vertical velocity $W^{(0)}(z)$.


Finally, from \eqref{eq:appendix_cont} the horizontal component vertical mode is also a plane wave
\begin{equation}\label{eq:u_form}
u(x,z,t) = \mathrm{Re}\left\{ U(z)\exp(i(k_x x-\omega t))\right\},
\end{equation}
and the modes $U^{(n)}(z)$ can be computed from the $W^{(n)}(z)$. The fundamental mode horizontal velocity component $U^{(0)}(z)$, in particular, relates to the vertical component $W^{(0)}(z)$ as
\begin{equation}
U^{(0)}{(z)} = \frac{1}{k_{x}^{(0)}}\frac{\textrm{d}W^{(0)}}{\textrm{d}z},
\end{equation}
and $u=U^{(0)}{(z)\sin(\omega t)}$ is what is forced at the inlet boundary on the left of the domain as detailed in \S\ref{sec:computational_approach}. The derivative $\textrm{d}W^{(0)}/\textrm{d}z$ is approximated from the discretized $\mathbf{w}^{(0)}$ using centered finite differences. The solution of the resulting linear system provides a discretized $\mathbf{u}^{(0)}$ used to reconstitute $U^{(0)}(z)$. In the main text, for the sake of simplicity, we drop the subscript and note $U(z)\equiv U^{(0)}(z)$. 

\thispagestyle{empty}

\clearpage

\renewcommand{\thefigure}{S\arabic{figure}}

\setcounter{figure}{0}    

\thispagestyle{empty}
\section{Pycnocline thickness variation for constant energy injected}

\vspace{1em}

We present here analogous results to the ones obtained in \S\ref{sec:effect_pycnocline}, when instead of verifying that the waves arriving at $x=L$ have same energy $E_{k,0}$, we inject waves with the same energy, matching the case $\delta=\SI{0.2}{m}$ presented in \S\ref{sec:Boluses}, without considering the evolution of the wave until it reaches the slope.  This approach may be more relevant for experimental studies as it may not be possible to measure directly how much energy is present at the breaking location.  This strategy results in comparatively less energy being injected at the broader pycnocline thicknesses and more energy injected for the narrow pycnoclines.  The resulting (non-monotonic) $S_b$ and $D_b$ trends with $\delta$ are presented in figure \ref{fig:figureS1}.

As in the case where energy is constant at the breaking location, the size of the bolus has a maximum value at an intermediate pycnocline thickness.  This peak occurs for a narrower thickness between $\delta=\SI{0.1}{m}$ and $\SI{0.15}{m}$, while the peak for the constant energy at the breaking location study is at $\delta=\SI{0.15}{m}$.  Due to the relatively higher energies for the narrower pycnoclines, the maximum bolus size is also larger than those observed in \S\ref{sec:effect_pycnocline}.  The distance traveled upslope also has a similar overall trend in that the displacement upslope increases as $\delta$ increases.  However, unlike the constant energy at the breaking point example, there is ultimately a maximum displacement upslope reached at $\delta=\SI{0.3}{m}$.  The shorter travel upslope for broader pycnoclines is due to the relatively lower energy in the generated bolus.

\vspace{0.5cm}


\noindent
\begin{minipage}{\linewidth}
\makebox[\linewidth]{
  \includegraphics[width=1.05\textwidth]{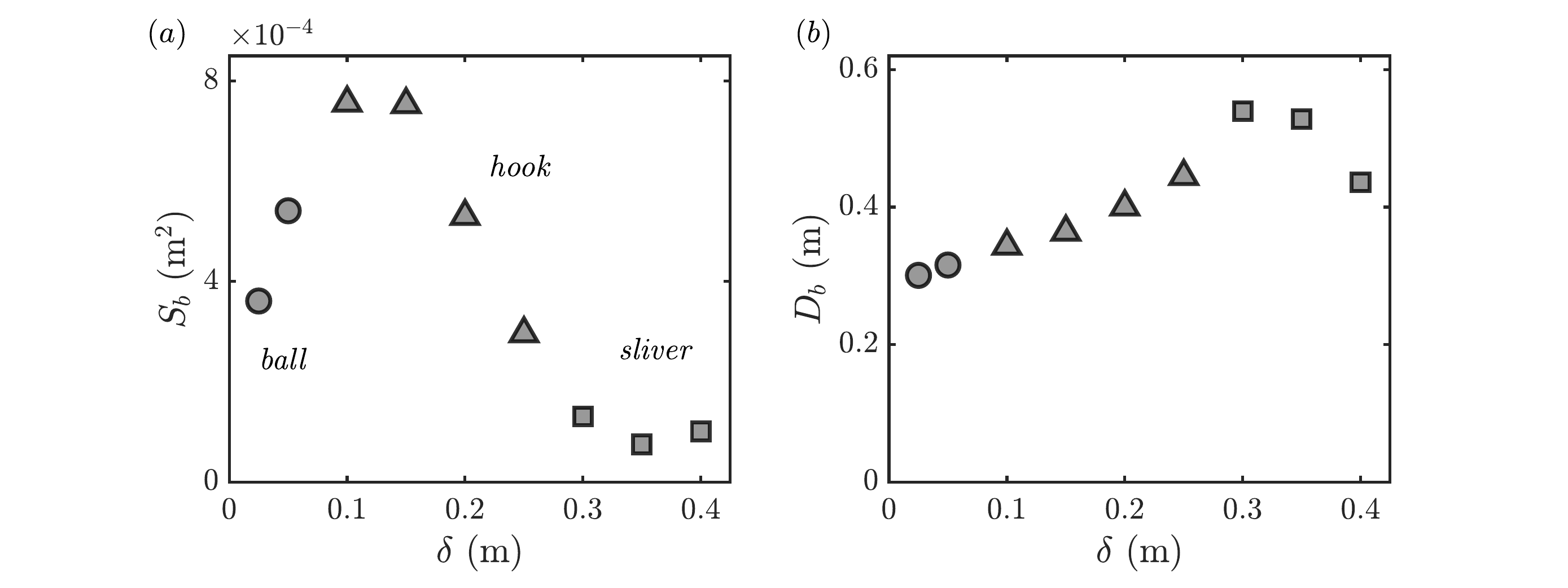}
  }
  \captionof{figure}{$(a)$ Bolus size $S_b$ and $(b)$ maximum displacement upslope $D_b$ as a function of the pycnocline thickness, $\delta$, for mode-1 waves breaking with constant energy. Different markers are used to represent bolus shape categories: ball (circle), hook (triangle) and sliver (square).}
  \label{fig:figureS1}
\end{minipage}

\clearpage

\renewcommand{\thefigure}{S\arabic{figure}}

\thispagestyle{empty}
\section{Wave characteristics and parameter variation}

\vspace{1em}

Here we present how the wave properties vary for the different studies presented in \S\S\ref{sec:effect_pycnocline}-\ref{sec:effect_slope}, in particular as the pycnocline thickness $\delta$ is varied. The incident wave amplitude $a$, wavelength $\lambda$ and wave speed $c_x$ are presented in figure~\ref{fig:figureR5}. Black markers are used for the varying $\delta$ study (\S\ref{sec:effect_pycnocline}), a yellow to red color map for the varying energy study (\S\ref{sec:effect_energy}) in figure~\ref{fig:figureR5}$(a$-$c)$, and a green to blue colormap for the varying $\D\rho$ study (\S\ref{sec:effect_drho}) in figure~\ref{fig:figureR5}$(d$-$e)$. The changes in the topographic slope do not impact the values considered here.

\vspace{0.5cm}


\noindent
\begin{minipage}{\linewidth}
\makebox[\linewidth]{
  \includegraphics[width=1.15\textwidth]{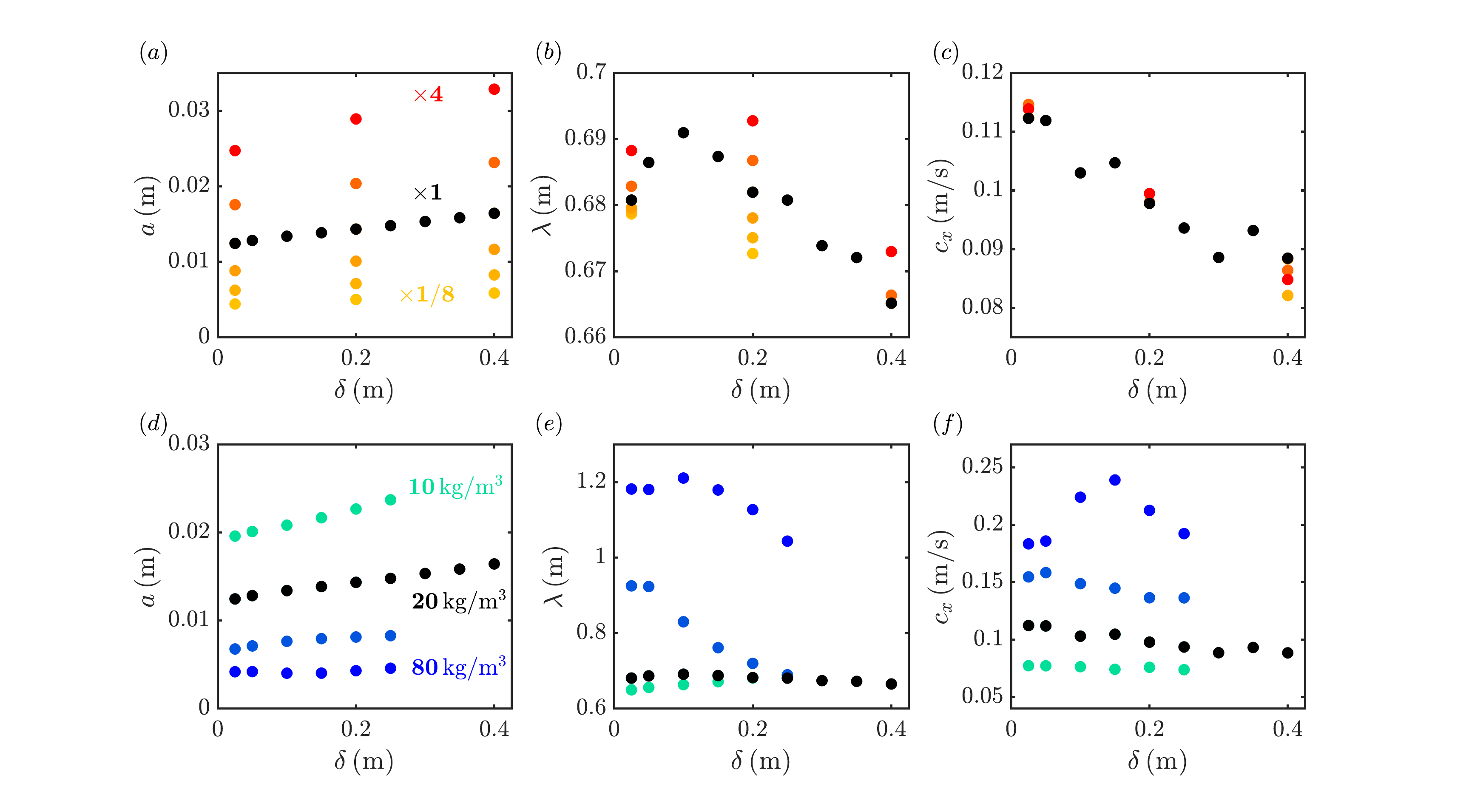}
  }
  \captionof{figure}{Amplitude $a$, wavelength $\lambda$ and propagation speed $c_x$ of the incident wave, as a function of $\delta$, for $(a$-$c)$ variable $E_k/E_{k,0}$, energy grows from yellow to red, $(d$-$f)$ variable $\D\rho$, density change grows from green to blue. Black markers represent the pycnocline thickness variation study, with $E_k/E_{k,0}=1$ and $\D\rho=\SI{20}{kg/m^3}$.}
  \label{fig:figureR5}
\end{minipage}

\clearpage

\renewcommand{\thetable}{S\arabic{table}}

\setcounter{table}{0}

\thispagestyle{empty}
\section{Internal wave beam and bolus transport}

\vspace{1em}

Internal waves generated by an oscillating tidal flow on a topographic slope were found to have highest amplitudes when the topographic slope matches the angle of propagation  of the internal wave beam $\theta$ (with the horizontal), such that $\sin \theta = \omega/N$~\citep{zhang08}.
If transport by boluses behaves similarly, it is reasonable to conjecture that internal waves shoaling on a constant slope topography will lead to maximum transport when the topographic slope $s$ equals the internal wave beam slope $s_\theta=\tan(\theta)$.  

Because $N$ varies vertically, so does $\theta$ and we are primarily concerned with the beam angle at mid-depth, $\theta_{H/2}$, where the boluses are generated. This beam angle depends on the magnitude of $\textrm{d}\rho_0/\textrm{d} z(H/2)$. If the topographic slope remains constant and transport is maximum at the internal wave critical angle, then as we vary $\D\rho$, maximum transport will happen in thinner pycnoclines for small $\D\rho$ and broader pycnoclines for large $\D\rho$.

However, as presented in table~\ref{tab:tab2}, $\theta_{H/2}$ for the stratifications producing the largest bolus for each $\Delta \rho$ differ from the topographic slope $s=0.176$, and the trends go against the conjecture that transport would maximize at the internal wave critical angle: boluses are larger for thinner pycnoclines when $\D\rho$ is larger and smaller for broader pycnoclines when $\D\rho$ is smaller.


\vspace{0.25cm}


\noindent\rule{\linewidth}{0.4pt}\vspace{1em}

\noindent
\begin{minipage}{\linewidth}
\makebox[\linewidth]{
\begin{tabular}{c@{\hskip 10pt}c@{\hskip 10pt}c@{\hskip 10pt}c@{\hskip 10pt}}
      $\D\rho\,(\SI{}{kg/m^3}$)  & $\delta\,(\SI{}{m}$)   &   $\theta_{H/2}$ & $s_\theta=\tan(\theta_{H/2})$ \\[5pt]
      10   &   0.2     &   \SI{41.5}{\degree} & 0.885\\
      20   &   0.15    &   \SI{24.0}{\degree} & 0.445\\
      40   &   0.1     &   \SI{13.6}{\degree} & 0.242\\
      80   &   0.1     &   \SI{9.5}{\degree}  & 0.167\\
  \end{tabular}
   }
\captionof{table}{Critical angle at mid-depth, $\theta_{H/2}=\theta(z=H/2)$, and corresponding critical slope $s_\theta$ for  parameter combinations $(\D\rho, \delta)_\textrm{max}$ that maximize the bolus size $S_b$.}
\label{tab:tab2}
\end{minipage}

\vspace{0.5em}

\noindent\rule{\linewidth}{0.4pt}

\end{document}